\documentclass{jfm}
\usepackage{graphicx}
\usepackage{amsmath,mathtools,amssymb,bm}
\graphicspath{{./fig/}}
\usepackage{epstopdf, epsfig}
\usepackage{color}

\newcommand{\dd}{{\rm d}}
\newcommand{\ii}{{\rm i}}
\newcommand{\beq}{\begin{equation}}
\newcommand{\eeq}{\end{equation}}
\newcommand{\ph}{\hat{\psi}}
\newcommand{\php}{\hat{\psi_p}} 
\newcommand{\phv}{\hat{\psi_r}} 
\newcommand{\Oh}{\hat{\Omega}}

\newcommand{\ccp}{C_\infty C_\infty^\prime}
\newcommand{\bc}{B C_\infty}
\newcommand{\bpcp}{B^\prime C_\infty ^\prime}
\newcommand{\Us}{U}
\definecolor{mygray}{rgb}{0.55, 0.488, 0.48}

\newcommand{\Pra}{\text{\textit{Pr}}}

\newcommand{\Schm}{\text{\textit{Sc}}}

\newcommand{\Reyn}{\text{\textit{Re}}}

\catcode`\@=11
\def\gtsim{\mathrel{\vcenter{\m@th\offinterlineskip
\hbox{$\hfill>\hfill$}\kern.5ex\hbox{$\hfill\sim\hfill$}}}}
\catcode`\@=12

\catcode`\@=11
\def\ltsim{\mathrel{\vcenter{\m@th\offinterlineskip
\hbox{$\hfill<\hfill$}\kern.5ex\hbox{$\hfill\sim\hfill$}}}}
\catcode`\@=12
\catcode`\@=12

\def\Xint#1{\mathchoice
	{\XXint\displaystyle\textstyle{#1}}%
	{\XXint\textstyle\scriptstyle{#1}}%
	{\XXint\scriptstyle\scriptscriptstyle{#1}}%
	{\XXint\scriptscriptstyle\scriptscriptstyle{#1}}%
	\!\int}
\def\XXint#1#2#3{{\setbox0=\hbox{$#1{#2#3}{\int}$}
		\vcenter{\hbox{$#2#3$}}\kern-.5\wd0}}

\def\dashint{\Xint-}

\shorttitle{Aerodynamics of planar counterflowing jets}
\shortauthor{A. D. Weiss, W. Coenen and A. L. S\'anchez}

\title{Aerodynamics of planar counterflowing jets}

\author{
      A. D. Weiss\aff{1}\corresp{\email{a2weiss@eng.ucsd.edu}},     %
          W. Coenen\aff{1},     %
   \and 
   A. L.  S\'anchez\aff{1}}

\affiliation{\aff{1}Department of Mechanical and Aerospace Engineering, 
					University of California San Diego,
					La Jolla, CA 92093--0411, USA}

\begin{document}

\maketitle

\begin{abstract}
The planar laminar flow resulting from the impingement of two gaseous jets of different density issuing into an open space from aligned steadily fed slot nozzles of semi-width $R$ separated a distance $2H$ is investigated by numerical and analytical methods, with specific consideration given to the high-Reynolds and low-Mach number conditions typically present in counterflow-flame experiments, for which the flow is nearly inviscid and incompressible. It is shown that introduction of a density-weighted vorticity-stream function formulation effectively reduces the problem to one involving two jets of equal density, thereby removing the vortex-sheet character of the interface separating the two jet streams. Besides the geometric parameter $H/R$, the solution depends only on the shape of the velocity profiles in the feed streams and on the jet momentum-flux ratio. While conformal mapping can be used to determine the potential solution corresponding to uniform velocity profiles, numerical integration is required in general to compute rotational flows, including those arising with Poiseuille velocity profiles, with simplified solutions found in the limits $H/R \ll 1$ and $H/R \gg 1$. The results are used to quantify the near-stagnation-point region, of interest in counterflow-flame studies, including the local value of the strain rate as well as the curvature of the separating interface and the variations of the strain rate away from the stagnation point. 

\end{abstract}

\begin{keywords}
\end{keywords}

\section{Introduction}
\label{section:Introduciton}

This study describes the impingement of two aligned gaseous jets of different density  counterflowing from opposed nozzles. Counterflow jets are ubiquitous in chemical engineering applications, including different variants with nozzle separations and feed conditions designed to optimize the specific mixing and reaction needs of the given application \citep{Tamir}. The closely related problem of a jet impinging on a flat surface is of utmost interest in connection with the aerodynamics of VTOL aircraft \citep{strand1960inviscid}. Recently, counterflow jets have found application in the field of biology for use in hydrodynamic stretching of DNA molecules \citep{renner2015stretching}. The specific conditions addressed here, namely, low-Mach-number jets with moderately large Reynolds numbers and nozzle separation distances of the order of the nozzle transverse size, are of interest in laminar counterflow burners, schematically represented in figure \ref{fig:sketchofproblem}, which are used in combustion experiments to characterize the response to strain of nonpremixed and premixed flames \citep{Petersbook}. Both axisymmetric and planar configurations are of interest in applications, with the former geometry analyzed in recent studies \citep{Bisetti,Jaime} and the  latter being investigated in the present paper by a combination of analytical and numerical methods.

Planar counterflowing jets have been subject to a number of studies, mostly for configurations with identical impinging jets, whose steady solution exhibits a symmetric structure closely related to the stagnation-point flow formed by a jet impinging on a flat wall. The latter problem has been investigated at length, to characterize both the heat transfer rate \citep{gardon1966heat,martin1977heat} and the resulting wall shear stress \citep{Phares}. An interesting subclass of these problems includes those in which the flow is inviscid, treated numerically for the case of a nozzle-free jet with different velocity profiles by \citet{Rubel} and \citet{Phares2}. Potential flow was analyzed using conformal mapping by \citet{levey1960back}, who examined the jet issuing from an aperture on a flat wall impacting normally on a parallel wall. The presence of a nozzle was described approximately for potential flow in the work of  \citet{strand1960inviscid} by prescribing the condition of parallel flow at a finite distance from the wall, thereby extending the classical result of an irrotational free jet impinging on a wall (or colliding against an identical free jet). \citet{li2011,li2013experimental} described the stability of the counterflowing jet flow with and without excitation to describe the oscillations of the stagnation plane. Effects of confinement on the oscillations were addressed in the stability analysis of \citet{pawlowski2006bifurcation} to characterize the dependence of the dynamics observed in experiments \citep{denshchikov1978interaction,denshchikov1983auto} on the Reynolds number and on the geometry, defined by the ratio of nozzle spacing to nozzle radius. 

Unequal counterflowing planar jets, with and without vorticity, have received considerably less attention. The previous studies have focused on colliding jets confined in channels \citep{gupta1996gas,hosseinalipour1997flow1,hosseinalipour1997flow2}, including the stability of the resulting configuration \citep{pawlowski2006bifurcation}. To the best of our knowledge, the case of two planar jets issuing into an open space from aligned nozzles, relevant for slot-jet counterflow combustors, has not been addressed in previous work.

\section{The flow structure in counterflow combustors}

\begin{figure}
	\centerline{\includegraphics[scale=1.0]{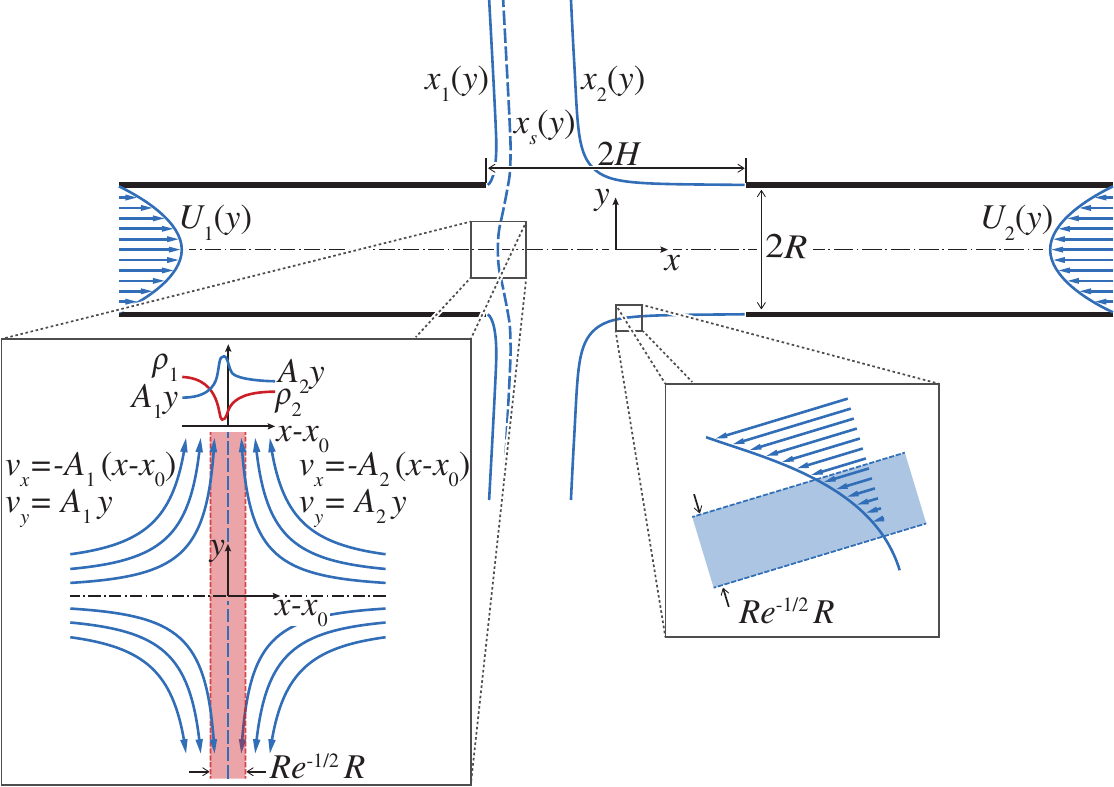}}
	\caption{Schematic representation of the counterflow configuration considered here, including a detailed view of the self-similar region around the stagnation point. The streamlines $x_1(y)$, $x_2(y)$, and $x_s(y)$ bounding the jets correspond to numerical integrations with $H/R=2$, $\rho^*_2/\rho^*_1=4$, $Q_2/Q_1=1.5$, $\Reyn=500$, and Poiseuille velocity profiles in the feed streams.}
	\label{fig:sketchofproblem}
\end{figure}

Counterflow burners are widely used in experiments of premixed, partially premixed, and non-premixed flames. The planar flow sketched in figure \ref{fig:sketchofproblem} is relevant in connection to slot-jet burners, used for instance in studies of edge-flame propagation \citep{ronney1,ronney2,ronney3,ronney4}. The spanwise length of these slot burners is sufficiently large to ensure that the resulting flow is locally planar away from the edges. In the figure, two opposed planar jets with volumetric flow rates $2Q_1$ and $2Q_2$ issue into a stagnant atmosphere from aligned screen-free nozzles of the same semi-width $R$ placed at a separation distance $2H$. In typical combustion experiments the mass fractions, density, temperature, and transport coefficients are uniform upstream from the nozzle exit, although they take in general different values in each of the feed streams, denoted by the subscripts $1$ and $2$, including densities $\rho_1^*$ and $\rho_2^*$ and viscosities $\mu_1^*$ and $\mu_2^*$. The shape of the velocity profiles in the feed streams depends on the development of the flow in the nozzle upstream from the exit plane. Sufficiently long nozzles result in Poiseuille velocity profiles, that being the case considered in figure~\ref{fig:sketchofproblem}, while short nozzles give velocity profiles that are uniform outside near-wall boundary layers. 

The mean jet velocity $U_m=Q_1/R \sim Q_2/R$ used in experiments is much smaller than the speed of sound, resulting in a low-Mach-number flow with spatial pressure variations that are much smaller than the ambient pressure. For laminar flame experiments, the specific selection of the geometry and injection conditions seeks to provide steady buoyancy-free laminar conditions in the central near-stagnation-point region, of primary interest for combustion tests (see \cite{Ulrich} for a detailed discussion of scaling criteria for counterflow burners). For instance, the jet velocity and nozzle size must be such that $U_m^2/(g R) \gg 1$ to minimize buoyancy effects. The typical values of the Reynolds number 
\beq
\Reyn = \frac{\rho^*_1 U_m R}{\mu^*_1} \sim \frac{\rho^*_2 U_m R}{\mu^*_2} \label{Reyn_def}
\eeq
range from about a hundred to about a thousand, so that the flow in the collision region is nearly inviscid. Molecular transport effects, including effects of viscous stresses, mixing, and heat conduction, are confined to thin layers, of small characteristic thickness $R/\Reyn^{1/2} \ll 1$. As indicated in the insets of figure \ref{fig:sketchofproblem}, one of the mixing layers is localized at the fluid surface separating the two jets, which departs from the central stagnation point, and the others at the fluid surfaces originating at the rims of the nozzles, separating the jets from the outer stagnant gas. For the moderately large values of the Reynolds number found in applications, the shear-driven instabilities affecting the mixing layers develop at a sufficiently slow rate for the central near-stagnation point region to remain virtually steady, as verified in recent direct numerical simulations \citep{Jaime}.

In combustion applications the flame is embedded in the mixing layer localized between the two opposing streams, where molecular transport and chemical reaction occur. The changes in temperature and density associated with the chemical heat release are confined to the interior of this thin separating mixing layer, whereas the temperature and density (and also the composition) remain uniform outside, and take in general different values on either side of the mixing layer, equal to those in the corresponding feed streams. Therefore, at leading order in the limit $\Reyn \gg 1$ the outer flow reduces to the inviscid collision of two jet streams of different density $\rho^*_1$ and $\rho^*_2$ bounded by sharp interfaces, whose location is to be calculated in a complicated free-boundary problem by using the condition of negligible pressure jump across the boundary interfaces \citep{Bergthorson}. The solution for the inviscid outer flow provides in particular the tangential velocity found on both sides of the interface separating the two jets, with a velocity jump occurring when the colliding jets have dissimilar density $\rho^*_1 \ne \rho^*_2$, as follows from the condition of equal pressure. This vortex-sheet character is not present in opposed jets with $\rho^*_1= \rho^*_2$, for which the resulting inviscid velocity field remains continuous at the separating interface, thereby simplifying the solution. 

The solution for the flow inside the slender mixing layer separating the two jets requires consideration of molecular-transport effects and, in combustion applications, also of chemical reactions. The problem, which can be formulated in the boundary-layer approximation, depends on the outer inviscid flow through the streamwise distribution of tangential velocity found on both sides of the mixing layer. The solution provides the transverse distributions of temperature, density, and composition, with boundary values given by the uniform properties of the bounding inviscid streams, different in general on both sides. As a result of the chemical heat release, the temperature is found to reach a maximum at the flame. Correspondingly, the density profile across the mixing layer, with boundary values $\rho^*_1$ and $\rho^*_2$, exhibits a minimum at the flame, as indicated in the inset of figure~\ref{fig:sketchofproblem}, whereas the radial velocity has an overshoot there, resulting from the action of the pressure gradient imposed on the heated gas by the outer flow. The solution for the inner structure of the mixing layer provides in particular the small values of the transverse velocity, of order $U_m/\Reyn^{1/2}$, found at the outer edges of the mixing layer, resulting from the thermal expansion associated with the chemical heat released at the flame. As shown by \cite{Kim}, these displacement velocities determine the first-order corrections to the outer inviscid flow, of order $\Reyn^{-1/2}$, which could be computed to increase the accuracy of the description, following a rigorous matched-asymptotic analysis for $\Reyn \gg 1$.

The description of the counterflow is simplified at distances from the stagnation point small compared with $R$, in a central region where the flow is self-similar, both outside and inside the mixing layer. The local velocity in the inviscid streams is given by the stagnation-point potential solution, including a radial velocity along the interface separating the two jets that increases linearly with the distance to the centreline $y^*$ according to $A^*_1 y^*$ in the inviscid stream 1 and $A^*_2 y^*$ in the inviscid stream 2. The stagnation-point strain rates $A^*_1$ and $A^*_2$, of order $U_m/R$, are related by $\rho^*_1 A_1^{* 2}=\rho^*_2 A_2^{* 2}$, as follows from the condition of equal pressure on both sides of the mixing layer. The flow in the reactive mixing layer in this near-stagnation-point region is also self-similar, with the temperature and composition varying with the distance to the stagnation plane. The resulting one-dimensional counterflow problem, which has been the basis for studies of flame-flow interactions in nonpremixed and premixed combustion \citep{Petersbook}, is amenable to numerical integration by standard commercial codes. The solution depends on the outer flow only through the value of the strain rate at the stagnation point $A_1^{*}=(\rho^*_2/\rho^*_1)^{1/2} A_2^{*}$, the reciprocal of which being the relevant local stretch time \citep{ARFM}. As mentioned above, at leading order in the limit $\Reyn \gg 1$ the value of $A_1^{*}=(\rho^*_2/\rho^*_1)^{1/2} A_2^{*}$ can be determined, with small relative errors of order $\Reyn^{-1/2} \ll 1$, from the analysis of the inviscid collision of two jets of different density, as done earlier in connection with axisymmetric counterflows \citep{Jaime}. The corresponding analysis for the case of planar jets, of direct interest for slot-jet counterflow burners, is to be presented below.

The description of inviscid counterflowing jets requires in general numerical integration of the Euler equations. It will be seen below in \S\ref{section:Formulation} that the problem can be reduced by introducing density-weighted variables to one involving equal densities in both streams, resulting in a continuous velocity across the surface separating both jets. Feed streams with uniform and Poiseuille velocity profiles will be considered in the analysis. The potential-flow solution associated with feed streams with uniform velocity profiles will be analyzed in \S\ref{section:Conformal} by conformal--mapping techniques based on Kirchhoff's method  \citep{birkhoff1957jets,gurevich1966,thomson1968theoretical}, with details of the needed mathematical development presented in an appendix. Numerical integrations of the Euler and Navier--Stokes equations will be used in \S\ref{section:Vortical} to quantify rotational flows for the case of Poiseuille velocity profiles in the feed streams. The analytical and numerical results are used to quantify many different relevant aspects of the flow, including the shape of the jet boundaries and the location and morphology of the near-stagnation-point region, the latter of direct interest in combustion applications. Finally, concluding remarks are given in  \S\ref{sec:conclusions}.

\section{Formulation}
\label{section:Formulation}


The problem to be analyzed below is the buoyancy-free low-Mach-number planar flow, symmetric about the centreline, resulting from the collision of two steadily fed gaseous jets of different density. Cartesian coordinates $\boldsymbol{x}=(x,y)$ centred at the middle point will be used in the description, with $x$ and $y$ denoting the longitudinal and transverse distances scaled with $R$. The mean velocity $U_m=Q_1/R$ in stream 1 will be used as characteristic scale for the dimensionless velocity $\boldsymbol{v}=(v_x,v_y)$, resulting in the nondimensional conservation equations 
\begin{eqnarray}
	\bnabla \bcdot (\rho\boldsymbol{v})&=&0,                           \label{con}  \\
	\rho \boldsymbol{v} \bcdot \bnabla \boldsymbol{v}&=&-\bnabla p' + \frac{1}{\Reyn} \bnabla \bcdot [\mu(\bnabla \boldsymbol{v} + \bnabla \boldsymbol{v}^T)], \label{mom}   \\
	\rho \boldsymbol{v} \bcdot \bnabla T&=& \frac{1}{\Pra \Reyn} \bnabla \bcdot (k \bnabla T),                                     \label{ener}   \\
	\rho \boldsymbol{v} \bcdot \bnabla Y_i &=& \frac{1}{\Schm_i \Reyn} \bnabla \bcdot (\rho D_i \bnabla Y_i),                    \label{Yeq}
	\end{eqnarray}
with the equation of state taking the simplified form
	\beq
	\rho T \sum_i \frac{W_1}{W_i} Y_i = 1.	\label{EqofState}
	\eeq
These equations must be complemented with expressions for the variation of the transport parameters $\mu$, $k$, and $D_i$ with the temperature and composition. The properties of stream 1 have been used to define dimensionless quantities, including the temperature $T$, density $\rho$, viscosity $\mu$, thermal conductivity $k$, and diffusion coefficient $D_i$ of chemical species $i$. The composition is described in terms of the mass fractions $Y_i$ of species $i$, with $W_i$ representing in~\eqref{EqofState} its molecular mass and $W_1=1/(\sum_i Y_{i1}/W_i)$ being the mean molecular mass in stream 1. In the momentum equation, $p'$ represents the pressure difference from the ambient value scaled with $\rho^*_1 U_m^2$. The parameters appearing in the above equations include the Prandtl and Schmidt numbers $\Pra$ and $\Schm_i$, and the Reynolds number $\Reyn$ defined in~\eqref{Reyn_def}, the latter taking moderately large values in typical combustion applications. 

\subsection{Nearly inviscid flow}

In the limit $\Reyn \gg 1$, the shear layers bounding the jets, of characteristic thickness $\Reyn^{-1/2} R$ at distances of order $R$, appear as infinitesimally thin surfaces. As indicated in figure~\ref{fig:sketchofproblem}, the jet issuing from the left nozzle is separated from the other jet by the streamline $x=x_s(y)$ departing from the stagnation point $\boldsymbol{x}_o=(x_o,0)$ and from the outer stagnant gas by the streamline $x=x_1(y)$ departing from the nozzle rim $(x,y)=(-H/R,1)$, with the streamline $x=x_2(y)$ departing from $(x,y)=(H/R,1)$ similarly separating the right-hand-side jet from the ambient atmosphere. These bounding surfaces are unknowns to be determined as part of the solution of a free-boundary problem, to be formulated below.

Consideration of the limit $\Reyn \gg 1$ in~\eqref{ener} and~\eqref{Yeq} leads to $\boldsymbol{v} \bcdot \bnabla T=\boldsymbol{v} \bcdot \bnabla Y_i=0$, indicating that the temperature and composition remain constant along any given streamline in the nearly inviscid jets outside the bounding mixing layers. Using these equations in~\eqref{EqofState} yields
\beq
\boldsymbol{v} \bcdot \bnabla \rho=0, \label{vrho}
\eeq
a result that can be used in~\eqref{con} to give 
\beq
\bnabla \bcdot \boldsymbol{v}=0 \label{solenoidal}
\eeq
and in~\eqref{mom} to give $\boldsymbol{v} \bcdot \bnabla (p'+\rho \boldsymbol{v}^2/2)=0$. The latter corresponds to the familiar condition of constant stagnation pressure $p'+\rho \boldsymbol{v}^2/2$ along streamlines, yielding in particular
\beq
p'+\rho \boldsymbol{v}^2/2=p'_o \label{pressurecenterline}
\eeq
along the centreline, where $p'_o$ is the pressure at the stagnation point $\boldsymbol{x}_o$, and 
 \beq
|\boldsymbol{v}|= \text{constant at } x=x_1(y) \text{ and at } x=x_2(y) \label{velpressure0}
\eeq
along the streamlines separating the jets from the stagnant air, where $p'=0$, with the constant taking in general different values on the surface of each jet.

According to~\eqref{vrho}, for inviscid incompressible flow each streamline carries the value of the density, so that the density field in the jet collision region is linked through~\eqref{vrho} to the upstream boundary distributions of density in the feed streams. Consequently, density variations across the feed streams result in a nonuniform density field in the jet collision region, determined by the curved streamlines, with associated density gradients that interact with the misaligned pressure gradients, generating vorticity, additional to that possibly carried by the feed streams. The analysis below is restricted to configurations involving feed streams with uniform temperature and uniform composition, typically encountered in experiments, for which the density in each one of the jets is uniform, given by $\rho=1$ in jet 1 and  $\rho=\rho^*_2/\rho^*_1$ in jet 2. Correspondingly, the baroclinic torque is negligible everywhere, except at the vortex sheet $x=x_s(y)$, where it generates a velocity jump according to
\beq
|\boldsymbol{v}|^2_{1}=(\rho^*_2/\rho^*_1) |\boldsymbol{v}|^2_{2} \text{ along } x=x_s(y), \label{veljump}
\eeq 
with the subscripts $1$ and $2$ denoting the velocities on the left and right sides of the interface. The above equation follows from the condition of zero pressure jump across the interface separating the two jets, with the pressure evaluated with use made of $p'+\rho \boldsymbol{v}^2/2=$ constant on both sides of $x=x_s(y)$. Along all other streamlines the vorticity magnitude $\omega=\partial v_y/\partial x-\partial v_x/\partial y$ remains constant, with a value equal to that found in the feed streams, to be determined from the boundary velocity distributions  
\beq \left\{
\begin{aligned}
	&\boldsymbol{v} = u_1(y)\boldsymbol{e}_x & {\rm as} \quad x \to -\infty, \\
	& \boldsymbol{v} =-({Q_2}/{Q_1})u_2(y)\boldsymbol{e}_x & {\rm as} \quad x \to \infty,
	\end{aligned} \right. \label{feedstream1}
	\eeq
where $u_1(y)$ and $u_2(y)$ represent nondimensional shape functions satisfying $\int_0^1 u_1 {\rm d}y=\int_0^1 u_2 {\rm d}y=1$. For example $u_1=u_2= 1$ and $u_1=u_2= (3/2)(1-y^2)$ for uniform and parabolic Poiseuille distributions, respectively. 


In the vicinity of $\boldsymbol{x}_o=(x_o,0)$, the velocity is given by the well-known stagnation-point distribution 
\beq \left\{
\begin{aligned}
& -v_x/(x-x_o)=v_y/y=A_1 & \text{in stream 1}, \\
& -v_x/(x-x_o)=v_y/y=A_2 & \text{in stream 2}, 
\end{aligned} \right. \label{potsol}
\eeq 
where the dimensionless strain rates $A_1=A_1^*/(Q_1/R^2)$ and $A_2=A_2^*/(Q_1/R^2)$ are related by
\beq
A_1=(\rho^*_2/\rho^*_1)^{1/2} A_2, \label{A1A2}
\eeq
consistent with~\eqref{veljump}. As previously mentioned, this inviscid value of the strain rate, to be quantified below in terms of the parameters defining the opposed-nozzle flow (i.e. $H/R$, $\rho^*_2/\rho^*_1$, and $Q_2/Q_1$), determines the near-stagnation-point solution for the thin reactive mixing layer separating the two jet streams. The accuracy of the selfsimilar mixing-layer solution, which applies strictly at the centreline, can be expected to degrade with increasing distances from the stagnation point as a result of curvature effects and of departures of the outer velocity from the linear distributions $v_y=A_1 y$ and $v_y=A_2 y$. Curvature effects can be quantified by computing the shape of the separating streamline $x=x_s(y)$ away from the centreline, given by
 \beq
x_s=x_o+y^2/(2r_c), \label{xsrc}
\eeq
with $r_c$ representing the local radius of curvature. On the other hand, the variations of the velocity on both sides of the separating vortex sheet are of the form
\beq
|\boldsymbol{v}|_1=(v_y)_1=A_1 y+A''_1 y^3/6 \text{ and } |\boldsymbol{v}|_2=(v_y)_2=A_2 y+A''_2 y^3/6, \label{veldists}
\eeq
written with account taken of the result $\p^2 v_y/\p y^2=0$ at $y=0$, stemming from the symmetry condition $\p v_x/\p y=0$ at $y=0$ and the solenoidal character of the velocity field. Clearly, configurations with smaller  values of $r_c^{-1}$ and $A''_1=(\rho^*_2/\rho^*_1)^{1/2} A''_2$ can be expected to exhibit an extended domain where the mixing layer is planar and is subject to a constant strain rate. In that respect, quantifications of the near-stagnation-point region based on the inviscid solution can be useful in assessing the range of validity of the one-dimensional selfsimilar description for the reactive mixing layer.

\subsection{Vorticity-stream function formulation}

In view of~\eqref{solenoidal} the problem can be formulated in terms of the standard stream function $\psi$, related to the vorticity by
\beq
    \frac{\partial^2 \psi}{\partial x^2} + \frac{\partial^2 \psi}{\partial y^2}	= -\omega.             \label{poisson}
	\eeq
The vorticity is constant along the streamlines, so that $\omega=\Omega(\psi)$, with the function $\Omega(\psi)$ determined implicitly through the expressions
	\beq \left\{
	\begin{aligned}
	\Omega &= -\frac{\dd u_1}{\dd y}, \quad \psi = \int_0^y u_1 \dd y, \quad \text{for} \quad 0\le \psi \le 1 \\
	\Omega&= \bigg(\frac{Q_2}{Q_1}\bigg)\frac{\dd u_2}{\dd y}, \quad \psi = -\bigg(\frac{Q_2}{Q_1}\bigg)\int_0^y u_1 \dd y,\quad \text{for} \quad -Q_2/Q_1 \le \psi \le 0
	\end{aligned} \right. \label{feeddist1} \eeq
derived with use of the boundary distributions \eqref{feedstream1}. The problem reduces to the integration of
\beq
    \frac{\partial^2 \psi}{\partial x^2} + \frac{\partial^2 \psi}{\partial y^2}	= -\Omega(\psi),             \label{poisson5}
	\eeq
supplemented with the implicit definition of $\Omega(\psi)$ given in~\eqref{feeddist1} and subject to the boundary conditions 
\begin{align}
\psi = 0 \; & \left\{ \begin{array}{l} \text{at} \; y = 0 \; \text{for} \; -\infty < x < + \infty \\
						\text{and at} \; x=x_s(y) \; \text{for} \; y\ge0, \end{array} \right. \label{bc1a} \\
\psi = 1 \; & \left\{ \begin{array}{l} \text{at} \; y = 1 \; \text{for} \; -\infty < x < - H/R \\
						\text{and at} \; x=x_1(y)   \; \text{for} \; y\ge 1, \end{array} \right. \label{bc1b} \\
\psi = -Q_2/Q_1 \; & \left\{ \begin{array}{l} \text{at} \; y = 1 \; \text{for} \; H/R < x < \infty \\
						\text{and at} \; x=x_2(y)   \; \text{for} \; y\ge 1. \end{array} \right. \label{bc1c} 
\end{align}
The surfaces $x_s(y)$, $x_1(y)$, and $x_2(y)$ are unknown free boundaries to be determined with use made of the additional boundary conditions~\eqref{velpressure0} and~\eqref{veljump}, written in the form
	\beq \left\{ \begin{aligned}
	&\tfrac{1}{2}|\bnabla \psi|^2 = p'_o - \tfrac{1}{2} [u_1^2(0) - u_1^2(1)] & \text{at} \quad x=x_1(y), \\
	&\tfrac{1}{2}({\rho^*_2}/{\rho^*_1})|\bnabla \psi|^2 = p'_o - \tfrac{1}{2}(\rho^*_2/\rho^*_1)(Q_2/Q_1)^2 [u_2^2(0) - u_2^2(1)] & \text{at} \quad x=x_2(y), 
	\end{aligned} \right. \label{velpressure1} \eeq
and
\beq
	|\bnabla \psi|^2_{1} = (\rho^*_2/\rho^*_1) |\bnabla \psi|^2_{2} \quad \text{at} \quad x=x_s(y), \label{seperatingsurface}
	\eeq	
respectively. In writing~\eqref{velpressure1} from~\eqref{velpressure0} we have used~\eqref{pressurecenterline} to express the pressure for the parallel flow in the nozzles far upstream from the exit in terms of the stagnation-point pressure $p'_o$.

\subsection{Reduction to the case of equal densities}

\label{sec:reducedformulation}

The problem defined in~\eqref{feeddist1}--\eqref{seperatingsurface} determines the stream function $\psi(x,y)$ along with the jet boundaries $x_1(y)$,  $x_2(y)$, and $x_s(y)$, and the stagnation-point pressure $p'_o$. Besides the shapes of the velocity profiles in the feed streams, defined by the functions $u_1(y)$ and $u_2(y)$, the solution depends on three parameters, namely, $H/R$, $Q_2/Q_1$, and $\rho^*_2/\rho^*_1$. As noted earlier in connection with axisymmetric jets \citep{Jaime}, the solution can be simplified by incorporating a renormalization factor $(\rho^*_2/\rho^*_1)^{1/2}$ in the definition of the kinematic variables for the fluid of density $\rho^*_2$. Specifically,  we introduce new density-weighted functions $\hat{\psi}$ and $\hat{\Omega}$, defined by 
\begin{equation}
\hat{\psi}= 
\begin{cases}
\psi, & \\
(\rho^*_2/\rho^*_1)^{1/2}\psi, &
\end{cases}
\text{   and    } 
\qquad \hat\Omega = \begin{cases}
\Omega, & \quad \text{for} \quad \psi>0 \\
(\rho^*_2/\rho^*_1)^{1/2}\Omega, & \quad \text{for} \quad \psi<0,
\end{cases} \label{defofph}
\end{equation}
to write~\eqref{poisson5} as
\beq
	\frac{\p^2 \ph}{\p x^2} + \frac{\p^2 \ph}{\p y^2} = -\Oh(\ph), \label{psihateqn}
	\eeq
	to be integrated with the boundary conditions
	\beq
	\begin{cases}
		\ph = 0 \quad &\text{at} \quad y = 0 \quad \text{for} \quad -\infty < x < +\infty \\
		\ph = 1 \quad &\text{at} \quad y = 1 \quad \text{for} \quad -\infty < x < -H/R \\
		\ph =-\Lambda \quad &\text{at} \quad y = 1 \quad \text{for} \quad H/R < x < +\infty \\
	\end{cases} \label{psihatbcs}
	\eeq
	and 
	\beq
	\begin{cases}
		\ph = 1, \quad & \tfrac{1}{2}|\bnabla \ph|^2 = p'_o - \tfrac{1}{2}[u_1^2(0) - u_1^2(1)], \quad \text{at} \; x=x_1(y)        \\
		\ph =-\Lambda, \quad & \tfrac{1}{2}|\bnabla \ph|^2 = p'_o - \tfrac{1}{2}\Lambda^2[u_2^2(0) - u_2^2(1)] \quad \text{at} \; x=x_2(y), \\
	\end{cases} \label{speed1}
	\eeq
which follow from~\eqref{bc1a}--\eqref{velpressure1}, respectively, whereas the dynamic condition~\eqref{seperatingsurface} is automatically satisfied provided that $\bnabla \ph$ is continuous, thereby removing the need to consider the separating surface $x_s(y)$ as a free boundary. In the reduced formulation the ratios $Q_2/Q_1$ and $\rho^*_2/\rho^*_1$ appear jointly in the new parameter 
\beq
\Lambda=\left(\frac{\rho^*_2}{\rho^*_1} \right)^{1/2} \left(\frac{Q_2}{Q_1} \right),
\eeq
with $\Lambda^2$ representing a measure of the ratio of jet momentum fluxes.  The function $\Oh(\ph)$, identically zero for uniform velocity in the feed streams, can be determined in general from~\eqref{feeddist1} written in the form
\beq \left\{
	\begin{aligned}
	\Oh &= -\frac{\dd u_1}{\dd y}, \quad \ph = \int_0^y u_1 \dd y, \quad \text{as} \quad x \to - \infty, \\
	\Oh &= \Lambda\frac{\dd u_1}{\dd y}, \quad \ph = -\Lambda\int_0^y u_2 \dd y, \quad \text{as} \quad x \to  \infty. 
	\end{aligned} \right. \label{inf1} \eeq
	 
Inspection of~\eqref{psihateqn}--\eqref{speed1} reveals that the transformation~\eqref{defofph} effectively simplifies the problem to one involving constant density, removing the vortex-sheet character of the separating surface $x_s(y)$, and concurrently reduces the number of controlling parameters from three to only two, namely, $H/R$ and $\Lambda$. Since the streamline pattern found with $\Lambda$ is a mirror image about the plane $x=0$ of that found with $1/\Lambda$, only flows with $\Lambda \ge 1$ need to be considered in the following. 

The jet outer interfaces $x_1(y)$ and $x_2(y)$ remain as unknown free boundaries, to be determined along with the unknown value of the stagnation pressure $p'_o$ as part of the computation of $\ph(x,y)$ for given values of $u_1(y)$, $u_2(y)$, $\Lambda$, and $H/R$. The solution provides the location of the stagnation point $\boldsymbol{x}_o=(x_o,0)$ and also the associated local values of
\beq
A_o=-\left.\frac{\p^2 \ph}{\p x \p y}\right|_{\boldsymbol{x}=\boldsymbol{x}_o}, \quad \frac{1}{r_c}=\frac{1}{3 A_o} \left.\frac{\p^3 \ph}{\p y^3}\right|_{\boldsymbol{x}=\boldsymbol{x}_o}, \; \text{and} \; A_o''=-\left.\frac{\p^4 \ph}{\p x \p y^3}\right|_{\boldsymbol{x}=\boldsymbol{x}_o}, \label{AorcAo''}
\eeq
with $A_o=A_1=(\rho^*_2/\rho^*_1)^{1/2} A_2$ and $A''_o=A''_1=(\rho^*_2/\rho^*_1)^{1/2} A''_2$. The free-boundary problem formulated above has analytical solutions only for configurations with uniform velocity distributions in the feed streams, such that $\Oh$ is identically zero, that being the case considered in the following section, with rotational solutions addressed in \S\ref{section:Vortical}.

\section{Irrotational counterflow jets}
\label{section:Conformal}

\subsection{General considerations}

When the velocity profiles in the feed streams are uniform, the vorticity function $\Oh$ in~\eqref{psihateqn} is identically zero, so that the inviscid flow is irrotational. With $u_1=u_2=1$, the stagnation pressure in both feed streams is uniform, equal to the stagnation-point pressure $p'_o$. The presence of the collision region generates an overpressure in the feed streams upstream from the nozzle exit, given by $p'_1=p'_o-1/2 \ge 0$ and $p'_2=p'_o-\Lambda^2/2 \ge 0$, respectively. The boundary condition \eqref{speed1} on the jet surface reduces to 
\beq \begin{cases}
		\ph = 1, \quad & \tfrac{1}{2}|\bnabla \ph|^2 = p'_o, \quad \text{at} \; x=x_1(y),        \\
		\ph =-\Lambda, \quad & \tfrac{1}{2}|\bnabla \ph|^2 = p'_o, \quad \text{at} \; x=x_2(y), \\
	\end{cases}  \eeq
revealing in particular that the speed $|\bnabla \ph|$ remains constant along these free surfaces, with a value given by $\Us=\sqrt{2 p'_o}$, to be determined as part of the analysis.

\begin{figure}
	\centerline{\includegraphics[scale = .825]{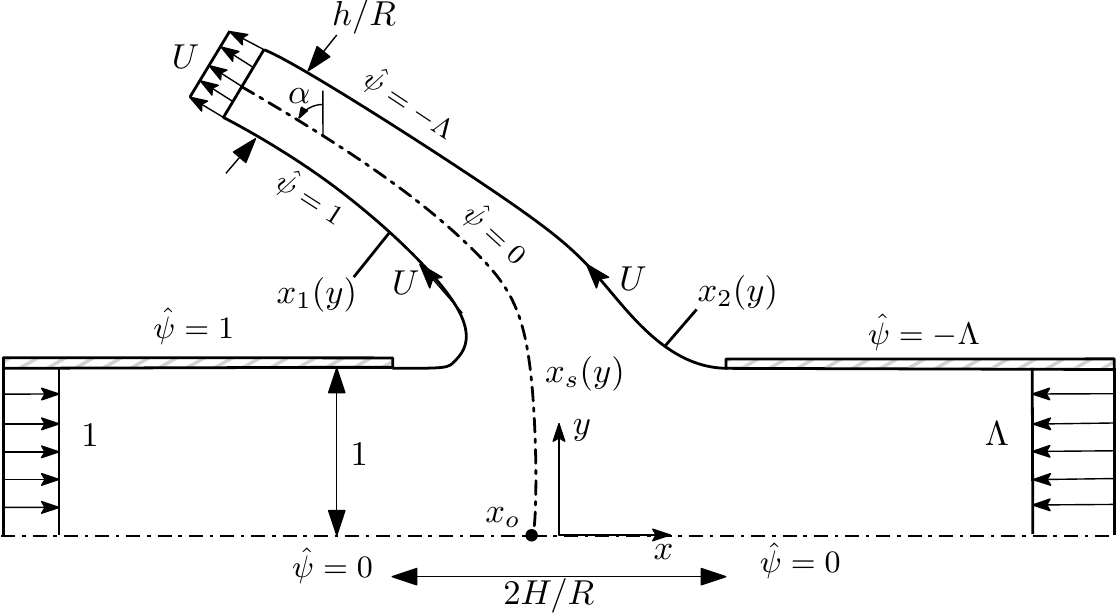}}
	\caption{Schematic view of the collision region for $H/R \sim 1$ and uniform velocity profiles in the feed streams.}
	\label{fig:complex_z_plane}
\end{figure}

The configuration depicted in figure~\ref{fig:complex_z_plane} corresponds to $H \sim R$, the case typically encountered in counterflow burners. Large and small values of $H/R$, also considered below, are of interest in chemical engineering applications \citep{Tamir}. For large separation distances, in the symmetric case $\Lambda = 1$ the problem reduces to the classical problem of collision of two free jets \citep{birkhoff1957jets,gurevich1966,thomson1968theoretical}, with the stagnation plane $x=0$ located far from both nozzle exits. This symmetric configuration is known to be prone to oscillatory instabilities \citep{li2011} that cause the stagnation plane to shift alternatively between both nozzle exits. Configurations with unbalanced momentum flux (i.e. $\Lambda \ne 1$) result in the collision region migrating to the vicinity of the nozzle carrying less momentum (the left nozzle in the cases $\Lambda >1$ considered here). In that case, the presence of a collision region has little effect on the flow at the outlet of nozzle 2, where the overpressure is $p'_2=0$ and the velocity profile remains uniform with magnitude $\Lambda$, so that for $H/R \gg 1$ and $\Lambda >1$ the constant free streamline speed $\Us$ is simply equal to $\Lambda$.

The collision of the opposed jets results in two symmetric jets that emerge laterally, as shown in figure~\ref{fig:complex_z_plane}. At transverse distances large compared with $R$ the streamlines in these lateral jets become aligned as the pressure approaches the ambient pressure $p'=0$, with the velocity correspondingly approaching the unknown uniform value $\Us=\sqrt{2 p'_o}$ across the jet, as follows from conservation of stagnation pressure along the streamlines. Conservation of mass and of longitudinal momentum provides the two relationships
\beq
\frac{h}{R} \Us=1+\Lambda \quad \text{ and }   \quad  \frac{h}{R} \Us^2 \sin \alpha=(\Lambda^2-1)/2, \label{balances}
\eeq
respectively, involving the thickness $h/R$ and deflection angle $\alpha$ of the lateral jet far from the opening, the latter measured relative from the $y$-axis. These two expressions can be combined with $\Us=\sqrt{2 p'_o}=\sqrt{\Lambda^2+2 p'_2}$ to give
\beq
\sin\alpha = \frac{\Lambda-1}{2\Us} = \frac{1}{2} \frac{1-1/\Lambda}{(1+2 p'_2/\Lambda^2)^{1/2}}, \label{sinalpha}
\eeq
involving the unknown overpressure $p'_2 \ge 0$ in the feed stream $2$. As can be seen, the value of $\alpha$, identically zero in the symmetric case $\Lambda=1$, increases for increasing values of $\Lambda$. An interesting conclusion stemming from~\eqref{sinalpha} is that for the opposed-jet arrangement investigated here (i.e. aligned jets issuing from nozzles of equal radius) the lateral-jet deflection is limited to a maximum value $\alpha=\upi/6$, achieved for $\Lambda \gg 1$ when the nozzles are placed far apart, so that the pressure in nozzle 2 equals the ambient value $p'_2=0$. 

It is worth noting that the potential description given here is limited to distances much smaller than $\Reyn R$, for which the thickness of the shear layers at the periphery of the lateral jets remain much smaller than the jet thickness $h \sim R$. Consideration of viscous effects, which are significant all across the jet at distances of order $\Reyn R$, would be needed to describe the transition of the jet velocity profile from the uniform value $\Us$ to the far-field Bickley profile, as done by \cite{virtual}.

\subsection{Selected potential-flow formulae}

\label{sec:PotentialHRorderUnity}

The solution simplifies for $H/R \ll 1$, in that the flow in the nozzles away from the small opening, including the central region near the stagnation point, is independent of the lateral emerging jets, with the opening acting as an apparent point sink of strength $(1+\Lambda)$ for the flow in the nozzles. The solution can be determined using superposition of a uniform flow and a sink in a channel \citep[see][]{thomson1968theoretical} to give
\begin{equation}
\ph=y-\frac{1+\Lambda}{\upi} \arctan\left[\frac{e^{\upi x} \sin(\upi y)}{e^{\upi x} \cos(\upi y)+1} \right]. \label{ph_potential_HR_0}
\end{equation}
The above expression can be used to determine the stagnation--point axial location from the condition $\p \ph/\p y=0$ at $y=0$ as well as the values of $A_o$, $r_c^{-1}$, and $A_o''$ from~\eqref{AorcAo''}, yielding
	\beq
	x_o = \frac{1}{\upi}\ln(1/\Lambda), \quad A_o = \frac{\upi \Lambda}{\Lambda + 1}, \quad r_c = \frac{3(\Lambda + 1)}{\upi(\Lambda-1)}, \; \text{and} \;  A''_o = -\frac{\upi^3 \Lambda(\Lambda^2-4 \Lambda+1)}{(\Lambda + 1)^3}, \label{xstagAspotentialHRzero}
	\eeq
independent of $H/R$. The high-speed flow at distances of order $H\ll R$ from the opening corresponds to the potential solution for the discharge of a pressurized container through an aperture, described for instance on pp.~310--311 of \cite{thomson1968theoretical}, including an emerging jet with
\beq
\frac{h}{R}=\frac{2\upi}{2+\upi} \frac{H}{R} \quad \text{and} \quad U=\frac{2+\upi}{2\upi} \frac{1+\Lambda}{H/R}, \label{hRUpotentialHRzero}
\eeq 
with the associated value of $h/(2H)=\upi/(2+\upi) \simeq 0.61$ being the well-known coefficient of contraction of a planar jet. Because of its large velocity, the deflection of the emerging jet, required to accommodate the unbalanced momentum flux of the opposed streams, is very small, as can be seen by using~\eqref{balances} and~\eqref{hRUpotentialHRzero} to write
\beq
\sin \alpha=\frac{\upi}{2+\upi} \frac{\Lambda-1}{\Lambda+1} \frac{H}{R}. \label{alphapotentialHRzero}
\eeq

The derivation of the corresponding analytic solution for the potential flow in the general case $H/R \sim 1$ requires use of conformal mapping techniques. A full and systematic discussion of the needed analysis is presented in Appendix~A. The development provides, in particular, the equation
\begin{align}
h/R =\cos\alpha \bigg\{2H/R + \frac{1}{\upi \Us^2}\bigg[&(U^2+1)\ln\bigg(\frac{U -1}{U+1}\bigg) + (U^2+\Lambda^2) \ln\bigg(\frac{U -\Lambda}{U+\Lambda}\bigg) \nonumber \\
&+ \frac{\Lambda^2-1}{2}\ln\bigg(\frac{2U+\Lambda-1}{2U-\Lambda+1}\bigg)      \bigg] \bigg\}, \label{deltax2_main}
\end{align}
relating the three unknowns $\alpha$, $h/R$, and $\Us$ with the parameters $\Lambda$ and $H/R$, as well as the expressions
\beq
\begin{aligned}
& A_o=\frac{\upi \Lambda U^4}{(U^2+\Lambda)^2(\Lambda+1)}, \; \frac{1}{r_c}=\frac{\upi U^2 (U^2-2 \Lambda)(\Lambda-1)}{3 (\Lambda+1)(U^2+\Lambda)^2}, \; A_o''=\frac{\upi^3 U^8 \Lambda}{(\Lambda+1)^3 (U^2+\Lambda)^6} \\
&\times [4U^2\Lambda (\Lambda-2) (2\Lambda-1) -U^4 (\Lambda^2-4\Lambda+1)-2 \Lambda^2 (3\Lambda^2-7\Lambda+3)]
\end{aligned}
\label{AoLambda_main}
\eeq
for the stagnation--point properties in terms of $U$ and $\Lambda$. Also, the conformal transformation provides integral expressions for the boundary surfaces $x_1(y)$ and $x_2(y)$, given in~\eqref{integralformoffreestreamlines1} and~\eqref{integralformoffreestreamlines2}, and for the separating streamline $x_s(y)$ departing from the stagnation point $\boldsymbol{x}_o=(x_o,0)$, obtained by integrating \eqref{dzofzeta} along the contour $\Gamma$ defined by~\eqref{contour}, with $x_o$ determined from~\eqref{xstageqn}.

Equations~\eqref{AoLambda_main}, valid in general for configurations with $H/R \sim 1$, enable quantification of stagnation-point properties in the limiting cases $H/R \ll 1$ and $H/R\gg 1$. For instance, the expressions given in \eqref{xstagAspotentialHRzero} for $A_o$, $r_c^{-1}$, and $A''_o$ when $H/R \ll 1$ may be recovered by letting $U \to \infty$ in~\eqref{AoLambda_main}, the appropriate limit as $H/R \to 0$. In the opposite limit $H/R \gg 1$ the free--stream velocity reduces to $U = \Lambda$, as discussed above, so that~\eqref{AoLambda_main} gives
\beq
A_o = \frac{\upi \Lambda^3}{(\Lambda + 1)^3}, \quad \frac{1}{r_c} = \frac{\upi \Lambda(\Lambda - 2)(\Lambda -1)}{3(1+\Lambda)^3}, \quad A''_o =-\frac{\upi^3(\Lambda -1)^2 \Lambda^5 (\Lambda^2-10\Lambda + 6)}{(1+\Lambda)^9}
\label{asympHRlarge}
\eeq
as the limiting values characterizing the stagnation point for distant nozzles with $H/R \gg 1$.

\subsection{General dependences on $H/R$ and $\Lambda$}

\begin{figure}
	\centerline{\includegraphics[scale = 0.5]{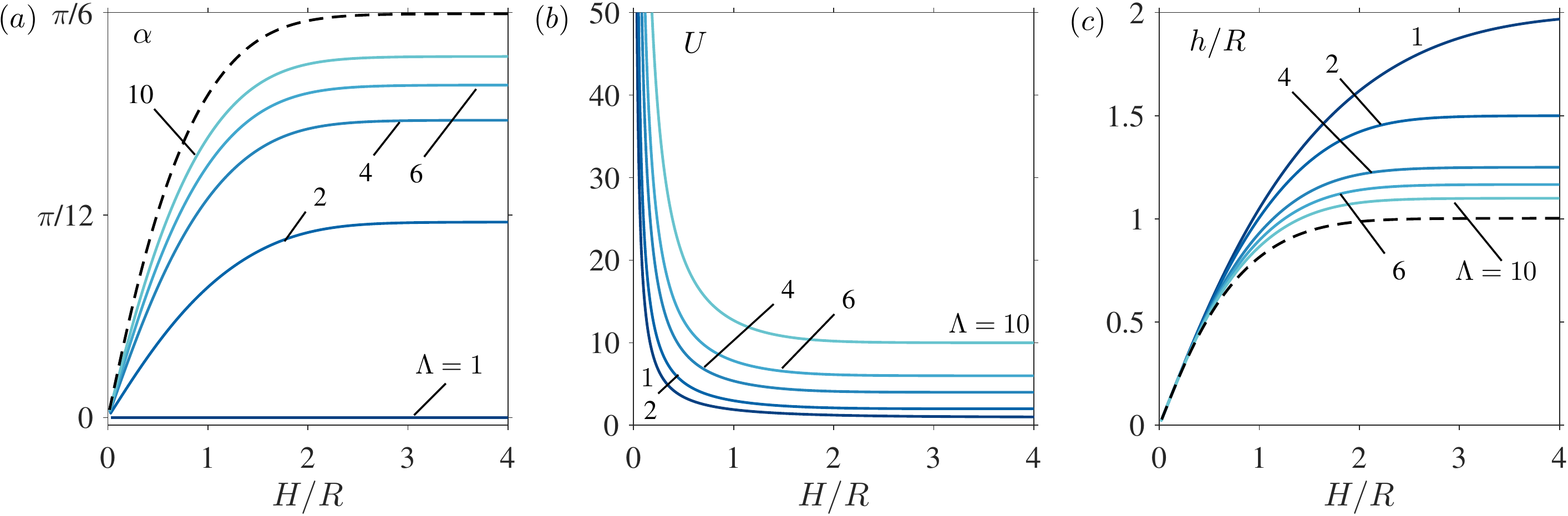}}
\caption{The variation with nozzle spacing $H/R$ of the deflection $\alpha$ of the emerging jet ($a$), surface speed $\Us$ ($b$), and emerging-jet width $h/R$ ($c$) for selected values of the momentum-flux ratio $\Lambda$. The dashed line in panels ($a$) and ($c$) correspond to $\Lambda \to \infty $.}
	\label{fig:stuff}
\end{figure}

The formulas developed above can be employed to determine the dependence of different flow features on the inter--nozzle spacing $H/R$ and on the momentum-flux ratio $\Lambda$. The free-surface velocity $U$ and the deflection angle $\alpha$ and dimensionless width $h/R$ of the emerging jet are computed by solving the mass and momentum conservation equations given in~\eqref{balances} together with~\eqref{deltax2_main}. For ease of calculation, it is convenient to select the values of $\Lambda $ and $\Us$ and then use \eqref{balances} to determine $\alpha$ and $h/R$, with the corresponding value of $H/R$ finally computed from \eqref{deltax2_main}. Results are shown in figure~\ref{fig:stuff} in the extended parametric ranges $0 < H/R \le 4$ and $1 \le \Lambda \le 10$.

The curves in figure~\ref{fig:stuff} help to quantify the effect of the inter--nozzle spacing $H/R$ on the resulting emerging jet. For small values of $H/R$ the inter--nozzle opening appears as a small gap, so that a large stagnation pressure $p'_o$ is needed to maintain the finite outflow rate $1+\Lambda$, resulting in emerging jets with larger velocity $U$ and smaller thickness $h/R$, as described by the asymptotic behaviors given in~\eqref{hRUpotentialHRzero} and~\eqref{alphapotentialHRzero}, corresponding to a jet discharging from a pressurized container through an aperture of width $2H$ on a flat wall. The opposite limit $H/R \gg 1$ corresponds to two nozzles placed at large distances. In the symmetric case $\Lambda=1$, the collision of the two jets, occurring at the middle point between the two distant nozzles, gives rise to two identical transverse jets of width $h=2R$ and velocity $U=1$, whereas with $\Lambda>1$ the collision region migrates to the vicinity of nozzle 1, where the outgoing stream encounters the opposed jet, which effectively behaves as a free jet with speed $U=\Lambda$. This is also the speed of the emerging jet for $H/R \gg 1$, whereas its thickness approaches $h/R=(1+\Lambda)/U=(1+\Lambda)/\Lambda$ in this same limit, as follows from continuity. The jet deflection, identically zero in the symmetric case $\Lambda=1$, is seen to increase with increasing $H/R$ for $\Lambda \ne 1$ to reach a maximum value $\alpha=\arcsin[(\Lambda-1)/(2\Lambda)]$ for $H/R \gg 1$, as can be obtained from~\eqref{sinalpha} with $U=\Lambda$. As pointed out earlier, the deflection is limited to a maximum value $\alpha=\upi/6$, reached as $\Lambda \rightarrow \infty$.

\begin{figure}
	\centerline{\includegraphics[scale = 0.6]{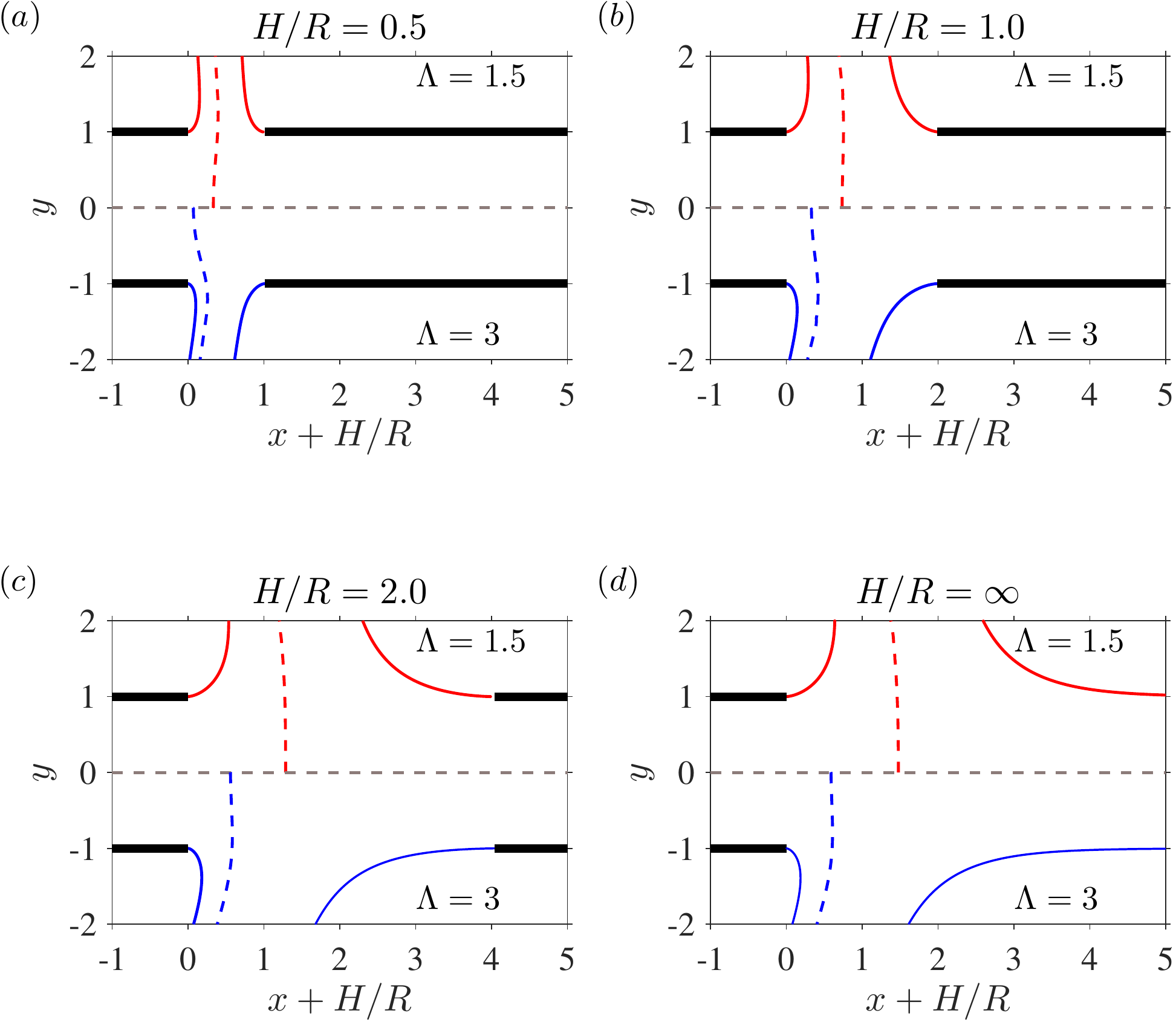}}
	\caption{Free--streamline patterns for different values of $\Lambda$ and $H/R$, including the outer interfaces $\ph=1$ and $\ph=\Lambda$ (solid curves) and the separating interface $\ph=0$ (dashed curves).}
	\label{fig:samplestreamlinepatterns}
\end{figure}

\begin{figure}
	\centerline{\includegraphics[scale = 0.6]{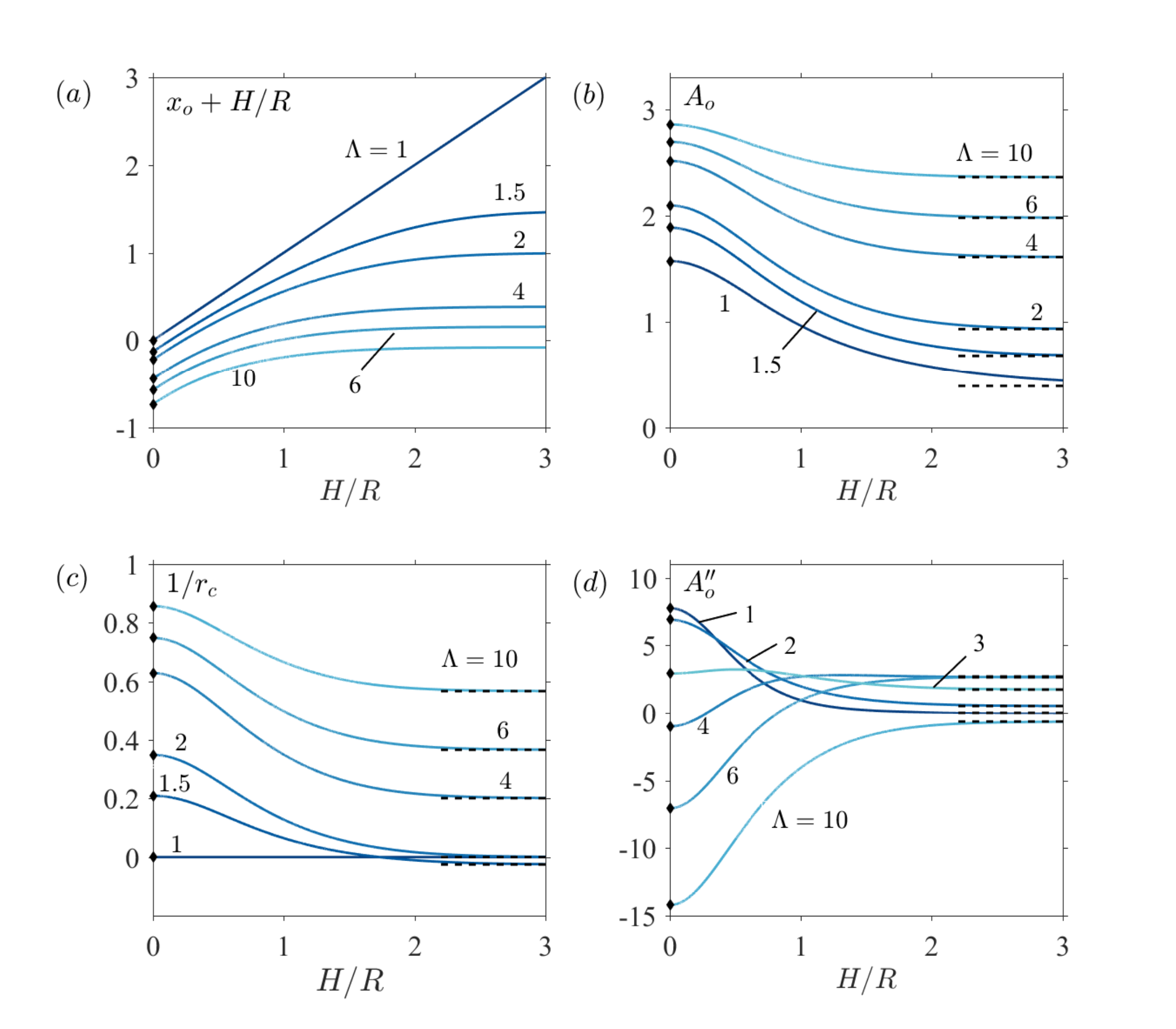}}
	\caption{The variation with $H/R$ of the distance of the stagnation point from the exit plane of the left nozzle $x_o + H/R$ ($a$), the stagnation-point strain rate $A_o$ ($b$), the local curvature of the separating interface at the stagnation point $1/r_c$ ($c$), and the parameter $A''_o$ measuring the departures of the velocity from the stagnation-point solution ($d$) for selected values of $\Lambda$; the dots along the vertical axes represent the limiting values given in~\eqref{xstagAspotentialHRzero} for $H/R=0$ whereas the dashed lines are the asymptotic values given in~\eqref{asympHRlarge} for $H/R \gg 1$.}
	\label{fig:strainrate_and_xstag_vs_HR}
\end{figure}

The interfaces $x=x_1(y)$, $x=x_2(y)$, and $x=x_s(y)$ are shown in figure~\ref{fig:samplestreamlinepatterns} for configurations with unbalanced momentum flux and different nozzle separation distances. As expected, a larger value of $\Lambda$ results in a displacement of the stagnation point towards the jet with smaller momentum and in a larger deflection of the emerging jet, the latter feature quantified in figure~\ref{fig:stuff}$a$. The variation with $H/R$ of the distance $x_o+H/R$ between the stagnation point and the left nozzle is shown in figure~\ref{fig:strainrate_and_xstag_vs_HR}$a$, with negative values of  $x_o+H/R$ corresponding to stagnation points lying inside the left nozzle. As can be seen, for a given momentum-flux ratio the stagnation point moves for increasing $H/R$ from the location $-\ln(\Lambda)/\upi$ corresponding to $H/R=0$ to reach a finite distance from the nozzle as $H/R \rightarrow \infty$, the only exception being the symmetric case $\Lambda=1$, for which the stagnation point keeps getting farther according to $x_o+H/R=H/R$. It is worth noting that for $\Lambda\gtsim 8.4$ the stagnation point remains inside the nozzle regardless of the value of $H/R$.

As previously mentioned, the flame in combustion experiments is embedded in the mixing layer separating the two jets, centred about the inviscid interface $x=x_s(y)$. Interest is focused on the near-stagnation-point region, where the solution for the inner structure of the reactive mixing layer is selfsimilar in the first approximation, determined by stagnation-point values of the strain rate $A_1$ and $A_2$ found in both colliding streams outside the mixing layer, related by $A_o=A_1=(\rho^*_2/\rho^*_1)^{1/2} A_2$. For slot nozzles with upstream uniform velocity profiles in the feed streams the analytic expression given in~\eqref{AoLambda_main} can be employed to determine $A_o$, with $\Lambda$ and $H/R$ entering as the only controlling parameters. The value of $A_o$ evaluated from~\eqref{AoLambda_main} with the surface velocity $U$ obtained from solving~\eqref{balances} and~\eqref{deltax2_main} is shown in figure~\ref{fig:strainrate_and_xstag_vs_HR}$b$ as a function of $H/R$ for different $\Lambda$. The results indicate that $A_o$ increases with increasing $\Lambda$ and with decreasing $H/R$, so that the resulting values range between the minimum $A_o=\upi/8$, corresponding to symmetric configurations with distant nozzles, to the maximum $A_o=\upi$, found irrespective of the nozzle spacing when the momentum flux is severely unbalanced. 

The selfsimilar solution for the counterflow mixing layer assumes a  locally planar structure, with negligible curvature. In reality, however, the separating interface is curved, and there is interest in quantifying this curvature to assess departures from locally planar selfsimilar structures. This is done in figure~\ref{fig:strainrate_and_xstag_vs_HR}$c$ with use made of~\eqref{AoLambda_main}. The plots reveal that the curvature is always positive for $\Lambda>2$, corresponding to convex interfaces. For values of the momentum-flux ratio in the range $1 < \Lambda < 2$, however, the curvature becomes negative for sufficiently large values of $H/R$, with a minimum value, reached as $H/R \rightarrow \infty$, given in~\eqref{asympHRlarge}. This change of sign of the curvature from convex to concave as the inter-nozzle distance is increased is apparent in the separating interfaces shown in figure~\ref{fig:samplestreamlinepatterns} for $\Lambda=1.5$. The results indicate that for configurations with $H/R \sim 1$ and $\Lambda \ltsim 2$, most often found in typical experimental arrangements, the separating streamline remains fairly flat, a finding that supports the neglect of curvature effects in analyzing the structure of the reactive mixing layer. 

Also central to the selfsimilar character of the counterflow mixing layer is the assumption that the velocity in the outer inviscid stream increases linearly with the streamwise distance, as occurs sufficiently close to the stagnation point. With increasing distances, however, the deviations from the linear acceleration become more noticeable, with the nondimensional parameters $A''_1$ and $A''_2$ measuring the extent of the departures, as indicated in~\eqref{veldists}. The value of $A''_o=A''_1=(\rho^*_2/\rho^*_1)^{1/2} A''_2$ evaluated from~\eqref{AoLambda_main} is shown in figure~\ref{fig:strainrate_and_xstag_vs_HR}$d$. For $H/R \ltsim 1$, the differences in values of $A''_o$ for different $\Lambda$ can be fairly large, with small values of $\Lambda$ resulting in accelerating flows with $A''_o>0$ whereas sufficiently large values yield decelerating velocities with negative $A''_o$. The differences diminish as the nozzle spacing increases. As seen in figures~\ref{fig:strainrate_and_xstag_vs_HR}$c$ and~\ref{fig:strainrate_and_xstag_vs_HR}$d$, configurations with $1 \le \Lambda \le 2$ tend to produce relatively small values of $A''_o$ and $1/r_c$ for $H/R \gtsim 1.5$, delineating an attractive parametric range for experimental designs aimed at minimizing departures of the mixing-layer structure from one-dimensional solutions.

\section{Counterflowing jets with distributed vorticity}
\label{section:Vortical}

We now investigate flows in which $\Oh \neq 0$, specifically considering the case of long nozzles with Poiseuille flow in each feed stream, for which $\Oh(\ph)$ is given by the implicit representation 
	\begin{align}
	\ph = 
	\begin{cases}
	(\Oh/6) \{3 - (\Oh/3)^2\}  \quad \text{for} \quad \ph >0\\
	(\Oh/6) \{3 - [\Oh/(3\Lambda)]^2\}\quad \text{for} \quad \ph <0
	\end{cases} \label{Omegahat_poise}
	\end{align}
as follows from~\eqref{inf1} with $u_1=u_2= (3/2)(1-y^2)$. The description of the inviscid flow requires numerical integration of~\eqref{psihateqn}--\eqref{speed1}, with the interfaces $x=x_1(y)$ and $x=x_2(y)$ entering as free boundaries. The constant speeds along these interfaces are given from~\eqref{speed1} by $U_1=[2 p'_o-(3/2)^2]^{1/2}$ and $U_2=[2 p'_o-(3/2)^2 \Lambda^2]^{1/2}$, respectively, where $p'_o$ is the stagnation-point pressure. The boundary distribution of stagnation pressure $p'+\tfrac{1}{2}|\bnabla \ph|^2$ is given by $p'_o+\tfrac{1}{2} (3/2)^2[(1-y^2)^2-1]$ in stream 1 and by $p'_o+\tfrac{1}{2} (3/2)^2 \Lambda^2 [(1-y^2)^2-1]$ in stream 2, respectively. Since the stagnation pressure is different for different streamlines, the velocity across the emerging jet exhibits a nonuniform distribution in the far field, where $p'=0$, with the maximum velocity, equal to $\sqrt{2 p'_o}$, found along the separating streamline $x=x_s(y)$.

The solution simplifies for $H/R \ll 1$ when the description of the flow in the nozzles away from the openings does not require consideration of the separating interfaces. This case is considered in~\S\ref{subsection:VorticalH/R<<1}, which includes comparisons with the irrotational results derived earlier in \S\ref{sec:PotentialHRorderUnity} to investigate influences of boundary velocity distributions. Next, we shall consider configurations with $H/R \sim 1$. Instead of solving the complicated free-boundary problem~\eqref{psihateqn}--\eqref{speed1} arising in the inviscid limit, the large-Reynolds-number flow will be described through integrations of the complete Navier--Stokes equations for $100\le \Reyn \le 500$. Results of a selected group of computations with uniform feed streams will be compared with the exact potential solution to validate the numerical description and also to test the reduced parametric dependence identified above. 

\subsection{Rotational flow with $H/R \ll 1$}
\label{subsection:VorticalH/R<<1}

\begin{figure}
	\centerline{\includegraphics[scale = 0.5]{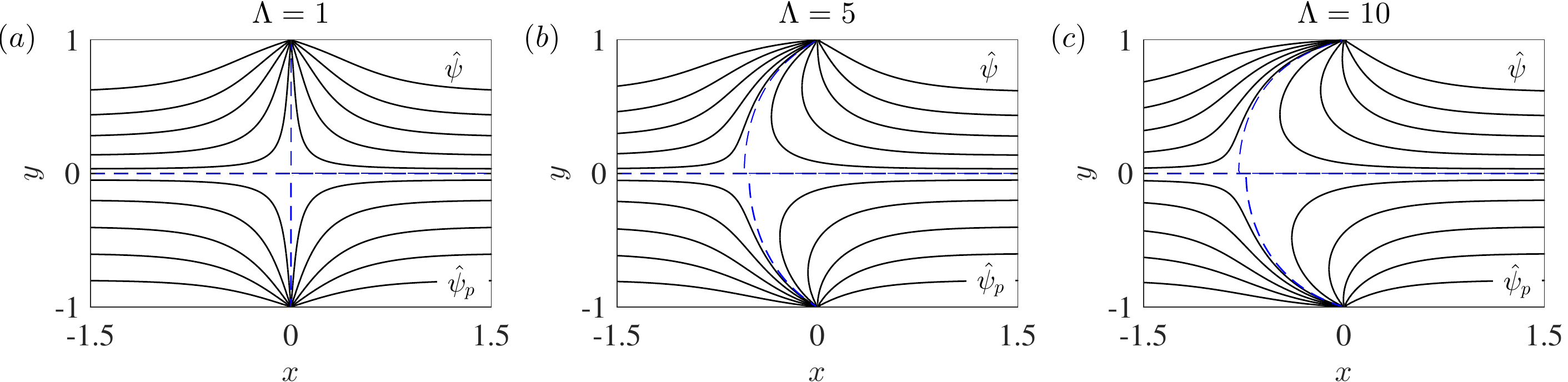}}
	\caption{Streamlines corresponding to $H/R=0$ for feed streams with Poiseuille velocity profiles (upper half of the plots) and with uniform velocity profiles (lower half of the plots). Besides the separating streamline $\ph=0$, denoted with a dashed curve, the plots show the streamlines $\ph=(0.05,0.2,0.4,0.6,0.8)$ for stream 1 and the streamlines $\ph=-\Lambda \times (0.05,0.2,0.4,0.6,0.8)$ for stream 2.}	
		\label{Vortical_HOR_0_streamlines}
\end{figure}

For cases with $H \ll R$, the velocity at distances of order $H$ from the opening becomes a factor $R/H$ larger than that found in the feed streams, as follows from a straightforward continuity balance. The associated point--sink singularities arising in the limit $H/R \rightarrow 0$ at the apparent opening locations $(x,y)=(0,\pm1)$ can be effectively handled in the numerical description by expressing $\ph$ as the sum of a potential stream function $\ph_p$ carrying the volume flux of the two streams and a rotational stream function $\phv$ with associated zero volume flux. The rotational stream function satisfies
\begin{equation}
\frac{\p^2 \phv}{\p x^2}+\frac{\p^2 \phv}{\p y^2}=-\Oh(\ph), \label{psihatvorteqn}
\end{equation}
where $\Oh(\ph)$ must be evaluated from~\eqref{Omegahat_poise} in terms of $\ph=\php+\phv$ with $\php$ given in \eqref{ph_potential_HR_0}. The boundary conditions for $\phv$ reduce to
\begin{equation}
\phv=0 \quad {\rm at} \quad y=0,1 \quad {\rm for} \quad -\infty < x < \infty  \label{homogenousbc}
\end{equation}
and
\beq
\phv = \frac{y}{2}(1-y^2) \quad \text{as} \quad x \to -\infty \qquad \text{and} \qquad \phv = - \Lambda \frac{y}{2}(1-y^2) \quad \text{as} \quad x \to \infty, \label{poise_infvort}
\eeq
for $0\le y \le 1$. The stream function approaches the Poiseuille distributions according to
\beq \left\{
\begin{aligned}
& \phv = \frac{y}{2}(1-y^2) + C_{-\infty} e^{\lambda_1 x} F_1(y) & {\rm as} \; x \rightarrow -\infty, \\
& \phv = - \Lambda  \frac{y}{2}(1-y^2) + C_{+\infty} e^{-\lambda_1 x} F_1(y) & {\rm as} \; x \rightarrow +\infty, 
 \end{aligned} \right.
 \label{psihatvortinfmod}
 \eeq
 where $\lambda_1=2.59$ and $F_1$ are the smallest eigenvalue and corresponding eigenfunction of the homogeneous problem
 \beq
 F_n^{\prime\prime} + \bigg[\frac{2}{1-y^2} + \lambda_n\bigg]F_n =0; \quad F(0) = F(1) = 0,
 \eeq
 obtained by linearizing~\eqref{psihatvorteqn} about the Poiseuille velocity distribution. The constant factors $C_{\pm\infty}$ in~\eqref{psihatvortinfmod} are to be determined as part of the numerical integration of~\eqref{psihatvorteqn}.

\begin{figure}
 	\centerline{\includegraphics[scale=0.5]{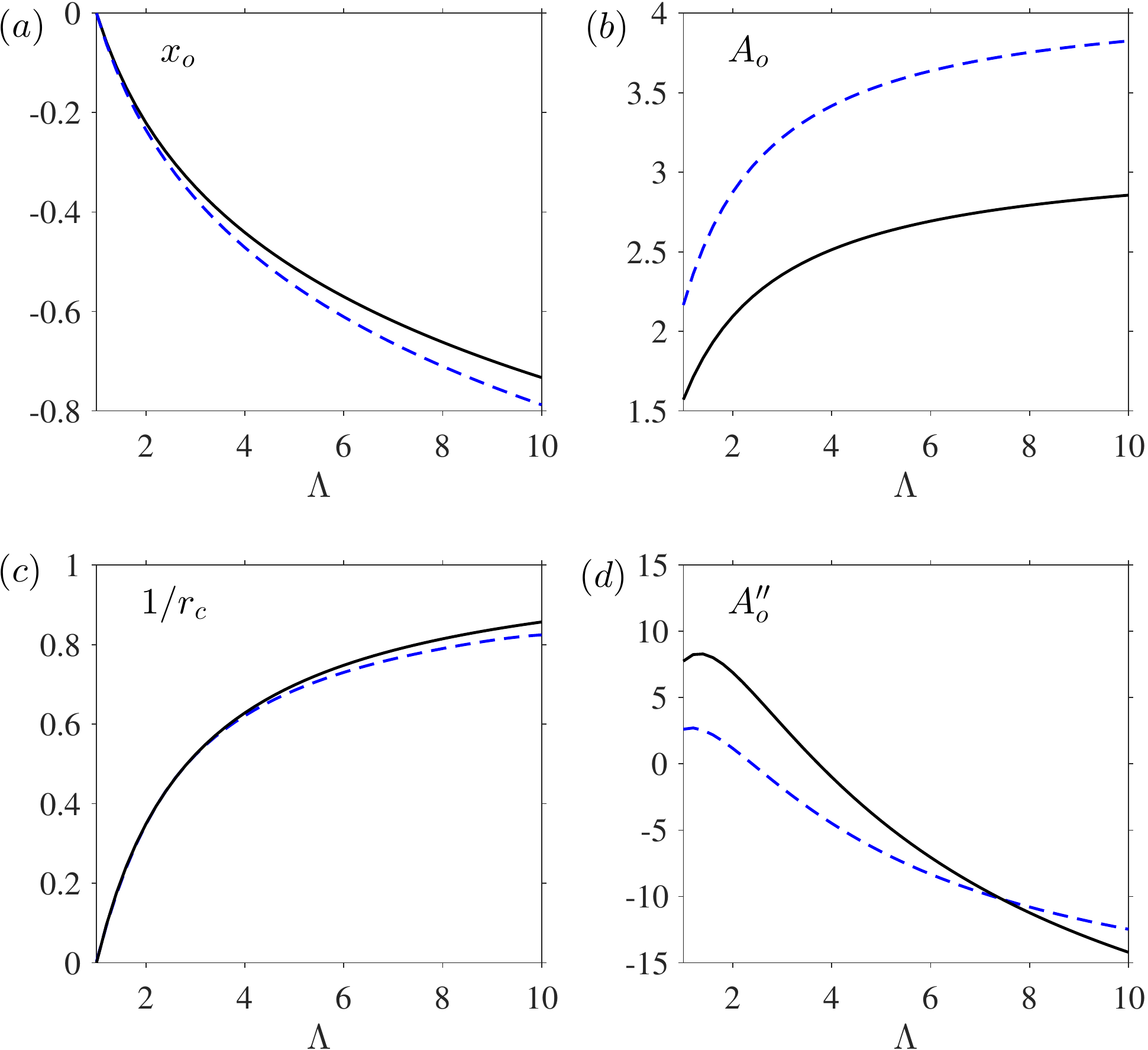}}
 	\caption{The variation with $\Lambda$ for $H/R=0$ of the distance of the stagnation point from the exit plane of the left nozzle $x_o$ ($a$), the stagnation-point strain rate $A_o$ ($b$), the local curvature of the separating interface at the stagnation point $1/r_c$ ($c$), and the parameter $A''_o$ measuring the departures of the velocity from the stagnation-point solution ($d$) as obtained numerically for selected values of $\Lambda$ with boundary Poiseuille velocity profiles (dashed curves) and evaluated from~\eqref{xstagAspotentialHRzero} for uniform velocity profiles (solid curves).}
 	\label{Vortical_HOR_0_StrainvsLam_XstagvsLam}
 \end{figure}

Equation \eqref{psihatvorteqn} was solved iteratively on a fixed rectangular domain until convergence was achieved, determined by the condition $||\ph_r^{k+1} - \ph_r^{k}||_{\text{max}}<10^{-6}$. A second-order central-difference scheme was used to discretize~\eqref{psihatvorteqn} written at each iteration $k$ in the form $\nabla^2 \ph_r^{k+1} = \Oh(\ph^k)$, which includes a nonlinear solve for $\Oh(\ph^k)$ as defined by the implicit representation \eqref{Omegahat_poise}. To facilitate convergence of the integrations in the finite domain $x_{-\infty} \le x \le x_{+\infty}$, the boundary conditions~\eqref{poise_infvort} as $x \rightarrow \pm \infty$ were replaced by
\beq
\left\{
\begin{aligned}
& \ph_r - \frac{1}{\lambda_1}\frac{\p \ph_r}{\p x} = \frac{y}{2}(1-y^2) & {\rm at} \; x=x_{-\infty},  \\
& \ph_r + \frac{1}{\lambda_1}\frac{\p \ph_r}{\p x} = -\Lambda\frac{y}{2}(1-y^2) & {\rm at} \; x=x_{+\infty}, 
\end{aligned} \right.
\label{psivortinfmod2}
\eeq
derived with account taken of the asymptotic behaviors~\eqref{psihatvortinfmod}. The solution was found to be independent of the longitudinal extent of the integration domain for sufficiently large values of $-x_{-\infty}$ and $x_{+\infty}$, with the results shown in figures~\ref{Vortical_HOR_0_streamlines} and~\ref{Vortical_HOR_0_StrainvsLam_XstagvsLam} corresponding to $-x_{-\infty}=x_{+\infty}=5$.

The streamlines $\ph=\php+\phv=$~constant are shown in figure~\ref{Vortical_HOR_0_streamlines} for different values of $\Lambda$ along with the corresponding potential-flow results, evaluated with use made of~\eqref{ph_potential_HR_0}. As can be seen, the general morphology of the flow is not critically affected by the shape of the boundary velocity distribution. The resulting changes in connection with the stagnation point are further quantified in figure~\ref{Vortical_HOR_0_StrainvsLam_XstagvsLam}. In the range $1 \le \Lambda \le 10$, the relative changes in axial location $x_o$ and in the local curvature $1/r_c$ of the separating streamline $x=x_o+y^2/(2r_c)$ are of the order of 5 \%. By way of contrast, the local velocity field in the vicinity of $\boldsymbol{x}_o=(x_o,0)$ is much more sensitive to the shape of the velocity in the feed streams, as seen in figures~\ref{Vortical_HOR_0_StrainvsLam_XstagvsLam}$b$ and~\ref{Vortical_HOR_0_StrainvsLam_XstagvsLam}$d$. The higher velocity of the Poiseuille distribution at the centreline produces a significantly higher value of $A_o$. The parameter $A_o''$, measuring the variation of the strain rate along the separating interface also exhibits pronounced differences. In particular, for $\Lambda \sim 1$ the rate of increase of the strain rate is considerably larger for potential flow.

\subsection{Vortical flow with $H/R \simeq \textit{O}(1)$}

\begin{figure}
	\centerline{\includegraphics[scale = 0.5]{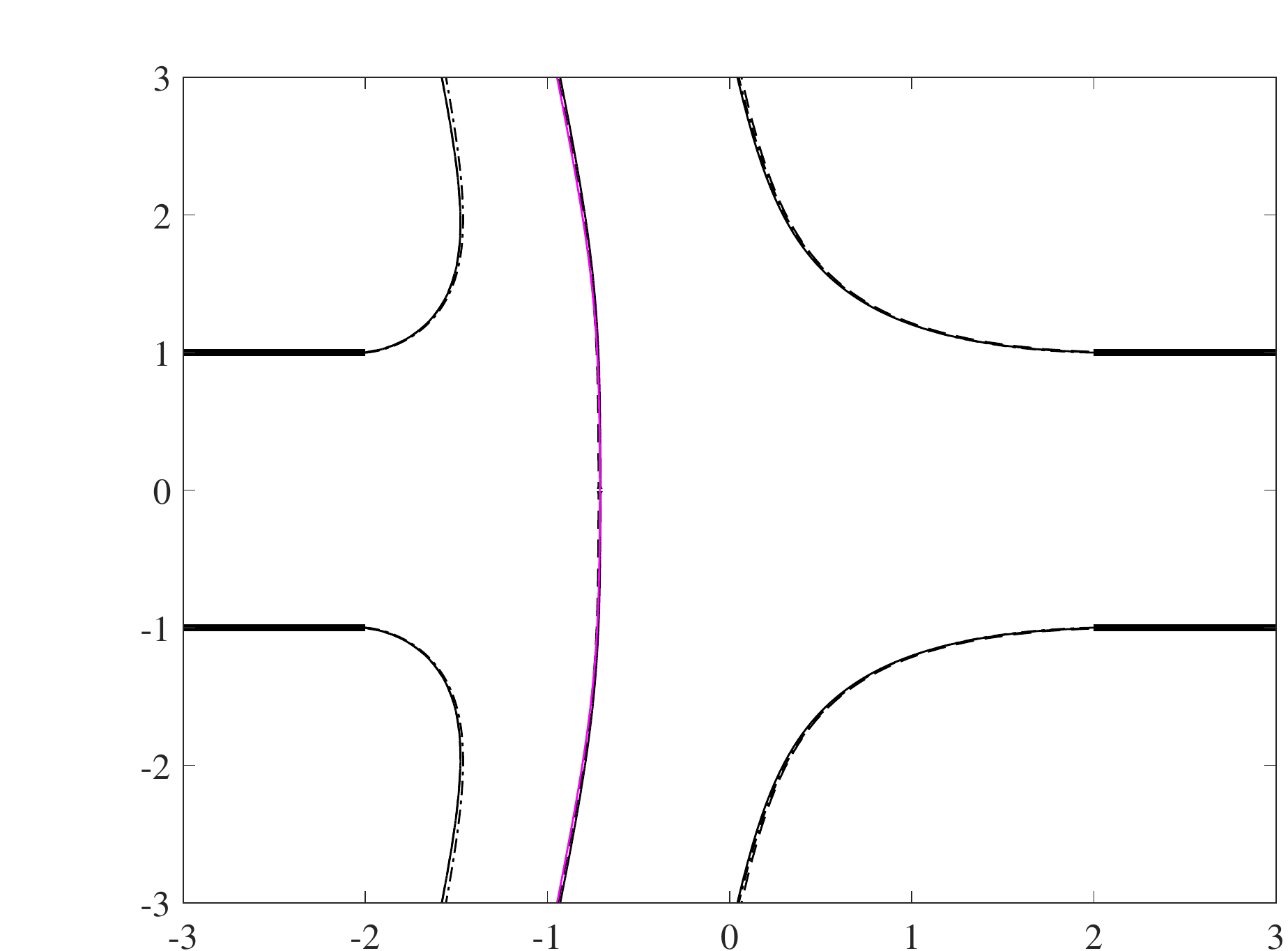}}
	\caption{Free and separating streamlines corresponding to uniform velocity in the feed streams and $H/R = 2$ obtained from numerical integrations of the Navier--Stokes equations for $\Reyn=500$ with $\rho^*_2/\rho^*_1 = 1$ and $Q_2/Q_1 = 3/2$ (solid curves), with $\rho^*_2/\rho^*_1 = 9/4$ and $Q_2/Q_1 = 1$ (dashed curves), and from evaluation of the potential--flow results of \S\ref{section:Conformal} (dash-dotted). All three cases correspond to $\Lambda=3/2$.}
	\label{fig:Validation}
\end{figure}

The computation of inviscid flows with free boundaries is in general a difficult task, especially with vorticity present. As shown by \citet{Bergthorson} when examining the related problem of an axisymmetric constant-density jet impinging on a perpendicular wall, the inviscid description can be approached by considering integrations of the Navier--Stokes equations for moderately large values of the Reynolds number, with associated departures from inviscid flow of order $\Reyn^{-1/2}$. This is the procedure followed below. 

The previous results concerning the parametric dependence of the problem in the limit $\Reyn \gg 1$ indicate that the specific selection of composition and temperature in the feed streams for the Navier--Stokes computations is not critical, in that different configurations with the same value of $\Lambda$ approach the same nearly inviscid flow structure in the limit $\Reyn \gg 1$, with differences arising only in the interior of the mixing layers surrounding the jets. For simplicity, the integrations below pertain to isothermal jets of two gases of different densities $\rho^*_2$ and $\rho^*_1$ but identical transport properties, such that $\mu=k=D_i=1$ and $\Schm = \Pra = 0.7$. The density of the ambient gas is taken to be equal to that of stream 1, so that the density at any point can be computed in terms of the mass fraction $Y_2$ by writing \eqref{EqofState} in the form $\rho=[1-Y_2 + Y_2(\rho^*_1/\rho^*_2)]^{-1}$, which can be used in~\eqref{Yeq} to give
\beq
\boldsymbol{v} \bcdot \bnabla \rho = \frac{\rho}{\Schm \Reyn} \bnabla \bcdot (\frac{1}{\rho} \bnabla \rho) \label{rhoeq}
\eeq
as a conservation equation for the gas density. The problem reduces to the integration of \eqref{con}, \eqref{mom}, and \eqref{rhoeq} with boundary conditions in the feed streams $v_x-(3/2)(1-y^2)=v_y=\rho-1=0$ as $x \to -\infty$ and $v_x+(3/2)(Q_2/Q_1)(1-y^2)=v_y=\rho-(\rho^*_2/\rho^*_1)=0$ as $x \to +\infty$. Nonpermeable walls with a nonslip flow condition are employed in the description, along with the symmetry conditions $\p v_x/\p y=v_y=\p \rho/\p y=0$ at $y=0$ and a condition of vanishing normal stress in the surrounding gas atmosphere, where $\rho=1$. The equations were integrated in the rectangular domain $x_{-\infty} \le x \le x_{+\infty}$ and $0\le y \le y_{+\infty}$, with $-x_{-\infty}=x_{+\infty} =3 H/R$ and $y_{+\infty}=7$. The integrations employed a finite-element method with $P1$ elements for the pressure field and $P2$ elements for the remaining variables, combined with a Newton-Raphson root-finding algorithm; details of the discretization method, used for instance by~\cite{Dani}, can be found in~\cite{Hecht_FF}. An illustration of the resulting flow is given in figure~\ref{fig:sketchofproblem}, where the boundary streamlines $x_1(y)$, $x_2(y)$, and $x_s(y)$ are obtained by integrating the above problem for $H/R=2$, $\rho^*_2/\rho^*_1=4$, $Q_2/Q_1=1.5$, $\Reyn=500$.

For the purpose of validation, a number of integrations were performed for uniform velocity in the feed streams, with a slip flow condition used at the wall to make the results of the integrations independent of the longitudinal extent of the integration domain. Figure \ref{fig:Validation} shows the separating streamlines (departing from the nozzle rims and from the stagnation point $\boldsymbol{x}_o$) obtained for $H/R=2$ from integrations of the Navier--Stokes equations with $\Reyn = 500$ for two different set of conditions, namely, $\rho^*_2/\rho^*_1-1=Q_2/Q_1-3/2=0$ and  $\rho^*_2/\rho^*_1-9/4=Q_2/Q_1-1=0$, resulting in the same momentum-flux parameter $\Lambda=3/2$. The associated curves are practically indistinguishable, supporting the reduced parametric dependence predicted by the density-weighted formulation presented in~\S\ref{sec:reducedformulation}. The figure includes comparisons with the potential--flow solution derived in \S\ref{section:Conformal}, whose associated jet interfaces $x=x_1(y)$, $x=x_2(y)$, and $x=x_s(y)$ are represented as dot-dashed curves. The close agreement of the potential flow with the Navier--Stokes results, expected in the limit $\Reyn \gg 1$ \citep{Bergthorson}, serves as validation for the numerical method.

\begin{figure}
\centerline{\includegraphics[scale = .7]{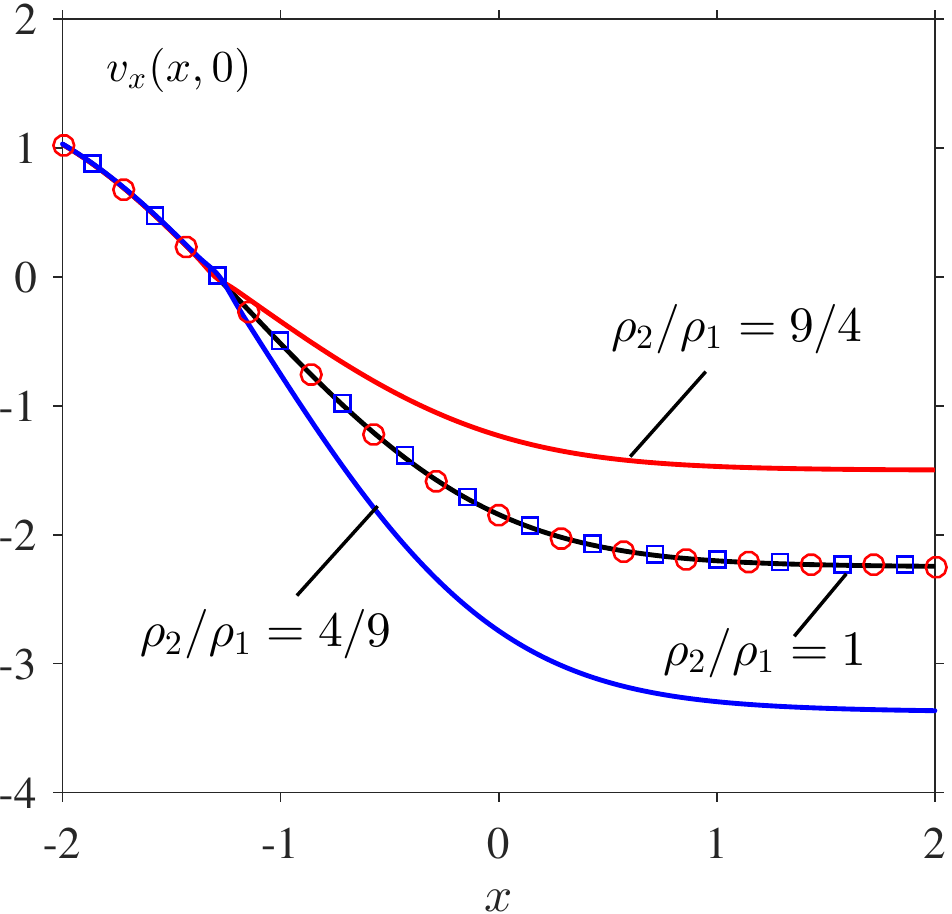}}
\caption{The distribution of longitudinal velocity along the centreline determined from the Navier--Stokes computations for $\Reyn=500$, $H/R=2$, and $Q_2/Q_1=1.5/(\rho^*_2/\rho^*_1)^{1/2}$ with $\rho^*_2/\rho^*_1=(4/9,1,9/4)$; the symbols represent the rescaled velocity distributions $\rho^{1/2} v_x$ for $\rho^*_2/\rho^*_1=4/9$ (squares) and for $\rho^*_2/\rho^*_1=9/4$ (circles).}	
\label{fig:comparee}
\end{figure}

To investigate the flow near the stagnation point the distribution of longitudinal velocity along the centreline obtained with Poiseuille velocity profiles in the feed streams is plotted in figure~\ref{fig:comparee} for $\Reyn=500$, $H/R=2$, and three different density ratios, with corresponding volume--flux ratios $Q_2/Q_1$ selected to maintain in all three cases the same value $\Lambda=(\rho^*_2/\rho^*_1)^{1/2} Q_2/Q_1=1.5$. The velocity on each side of the stagnation point outside the mixing layer follows a linear distribution given by $v_x=-A_1 (x-x_o)$ in stream 1 and  $v_x=-A_2 (x-x_o)$ in stream 2, in agreement with the local potential solution~\eqref{potsol}. The values of $A_1$ and $A_2$, which can be obtained by extrapolating the numerical results near the stagnation point, are seen to satisfy the relationship $A_o=A_1=(\rho^*_2/\rho^*_1)^{1/2} A_2$ stated in~\eqref{A1A2}. The resulting values, $A_o=(1.58,1.59,1.60)$ for $\rho^*_2/\rho^*_1=(4/9,1,9/4)$, are almost identical in all three cases, in agreement with the reduced inviscid formulation introduced earlier, which predicts that the strain rate $A_1=(\rho^*_2/\rho^*_1)^{1/2} A_2$ depends on $\rho^*_2/\rho^*_1$ and $Q_2/Q_1$ through the single parameter $\Lambda=(\rho^*_2/\rho^*_1)^{1/2}(Q_2/Q_1)$. The derivation of this reduced dependence is based on the density-weighted variables defined in~\eqref{defofph}, which are tested in figure~\ref{fig:comparee} by representing with symbols the distribution of $\rho^{1/2} v_x$ for the two cases with unequal jet densities. The resulting curves fall on top of the velocity distribution of the uniform-density case, further illustrating the applicability of the reduced inviscid formulation.

The dependence of the results on the Reynolds number was found to be relatively weak for moderately large values of $\Reyn$. Only small variations of a few percent were found in values of the stagnation-point location $x_o$ and in the stagnation-point values of $A_o$ and $A''_o$ for $100 \le \Reyn \le 500$, with somewhat larger variations affecting the curvature of the separating streamline $1/r_c$. For instance, for $H/R=2$, $\rho^*_2/\rho^*_1=1$, and $Q_2/Q_1=1.5$ the integrations provide $x_o=(-1.27,-1.29,-1.29)$, $A_o=(1.55,1.58,1.59)$, $A''_o=(1.07,1.07,1.07)$, and $1/r_c=(0.044,0.0561,0.0589)$  for $\Reyn=(100,300,500)$, respectively. 

\begin{figure}
	\centerline{\includegraphics[scale = .5]{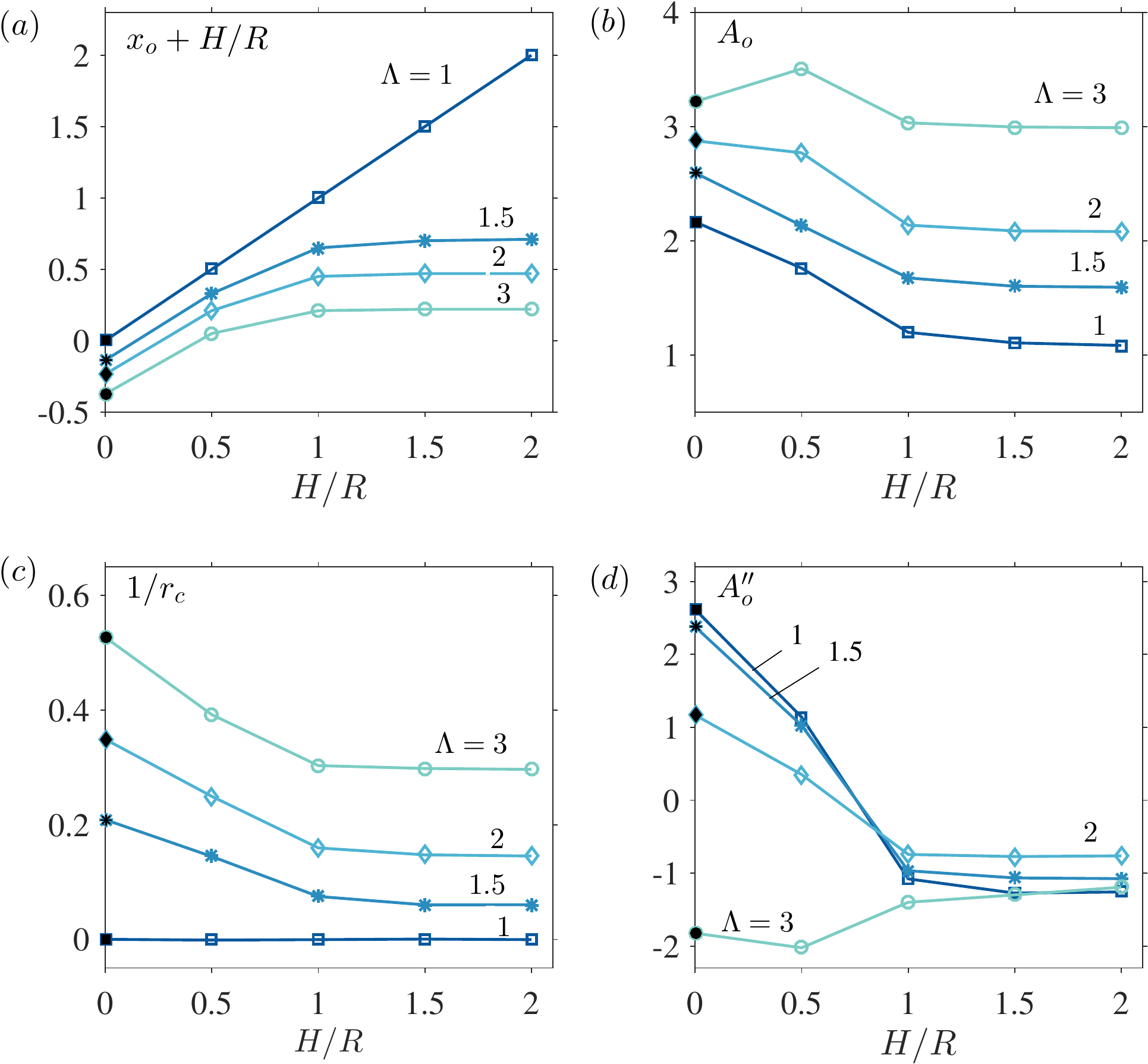}}
	\caption{The variation with $H/R$ of the distance of the stagnation point from the exit plane of the left nozzle $x_o + H/R$ ($a$), the stagnation-point strain rate $A_o$ ($b$), the local curvature of the separating interface at the stagnation point $1/r_c$ ($c$), and the parameter $A''_o$ measuring the departures of the velocity from the stagnation-point solution ($d$) obtained from the Navier--Stokes computations for $\Reyn=500$ and $\rho^*_2/\rho^*_1=1$; the solid symbols along the vertical axes represent the limiting values given in figure~\ref{Vortical_HOR_0_StrainvsLam_XstagvsLam}.}
	\label{fig:Vortical:StrainandStagvsHoR}
\end{figure}

The computations for $\Reyn=500$ and $\rho^*_2/\rho^*_1=1$ are used in figure~\ref{fig:Vortical:StrainandStagvsHoR} as a basis to investigate the dependence of $x_o$, $A_o$, $1/r_c$, and $A''_o$ on $H/R$ and $\Lambda$. The numerical evaluation of $A_o$ involves a linear fit about $x=x_o$ of the longitudinal velocity distribution $v_x(x,0)$. Similarly, $A''_o$ is obtained by computing the slope of a linear fit of the longitudinal distribution of $\p^2 v_x/\p y^2(x,0)$ about $x=x_o$, while the computation of $r_c$ requires a parabolic fit of the form~\eqref{xsrc} for the separating streamline departing from $\boldsymbol{x}_o$. The resulting curves, which include the limiting results for $H/R=0$ computed earlier in~\S\ref{subsection:VorticalH/R<<1}, exhibit trends that are qualitatively similar to those of the irrotational results in figure~\ref{fig:strainrate_and_xstag_vs_HR}.

\section{Conclusions}
\label{sec:conclusions}

The present paper contributes to the quantitative description of the steady planar flow resulting from the impingement of opposed low-Mach-number gaseous jets of different density issuing into an open stagnant atmosphere from aligned nozzles, with specific attention given to the high-Reynolds-number conditions prevailing in most chemical reactors \citep{Tamir} as well as in combustion facilities used to investigate the response of flames to strain \citep{Ulrich}. The flow is nearly inviscid in the outer streams, with effects of viscous forces, heat conduction, and species diffusion confined to thin mixing layers, one separating the two jets and the others separating each jet from the surrounding stagnant atmosphere. In combustion applications, the flame is embedded in the mixing layer separating the two jets, across which we find large density changes due to the heat released by the chemical reaction, whereas outside the density in each jet remains equal to the value found in its feed stream. The solution presented here for the inviscid collision of the two jets provides, amongst other things, the strain rate exerted by the outer flow on the separating mixing layer, a quantity of fundamental interest for studies of flame response to strain. 

A formulation involving a density-weighted stream function is introduced to simplify the problem to one involving two jets of equal density. The proposed formulation is also useful in reducing the number of controlling parameters to only two, namely, the dimensionless nozzle spacing $H/R$ and the ratio of jet momentum fluxes $\Lambda^2$. For uniform velocity in the feed streams, the resulting potential flow can be described analytically by conformal--mapping techniques, while numerical integration is needed to describe rotational flow, which is analyzed here for the important case of Poiseuille flow in the feed streams. The theoretical and numerical analyses provide relevant quantitative information regarding the parametric dependence of the flow, including analytic expressions derived for potential flow, given for instance in~\eqref{balances}, \eqref{deltax2_main}, and \eqref{AoLambda_main}, which enable relevant flow characteristics to be readily evaluated for nozzle-flow types of laminar counterflows with thin near-wall boundary layers. 

The variation of the shape of the interfaces bounding the jets is investigated in figure~\ref{fig:samplestreamlinepatterns}, while the far-field characteristics of the associated lateral jet is shown in figure~\ref{fig:stuff}. The jet deflection $\alpha$ from the perpendicular direction, small when the two nozzles are placed at a small distance, is seen to increase with increasing inter-nozzle distances. As anticipated in \eqref{sinalpha} on the basis of a simple jet-momentum balance, the deflection is limited to a maximum value $\alpha=\upi/6$, reached when $H/R \gg 1$ and $\Lambda \gg 1$. Because of their interest in combustion applications, specific attention has been given to the morphology and velocity field of the near-stagnation-point region, with results summarized in figures~\ref{fig:strainrate_and_xstag_vs_HR} and~\ref{fig:Vortical:StrainandStagvsHoR}. The results reveal qualitative similarity in the stagnation region properties between irrotational flows and those with distributed vorticity, main differences entering in the larger values of the strain rate in the latter while in the former the flow accelerates away from the stagnation region at a faster rate. An attractive parametric range --- characterized by momentum flux ratios between $1\le \Lambda^2 \le 4$ and nozzle spacings $H/R \gtsim 1.5$ --- was identified aimed at minimizing departures of the mixing-layer structure from one-dimensional selfsimilar solutions.

The analysis helps to clarify the flow characteristics of planar open-duct and nozzle-flow types of counterflows, found in chemical reactors and counterflow burners. The associated quantitative information may aid in considerations of experimental designs of such devices.

\section*{Acknowledgements}
The inputs of Profs. S. Llewellyn Smith, J. Carpio, A. Li\~n\'an, and F.A. Williams on different aspects of this research are gratefully acknowledged. 

\appendix
\section{Conformal mapping solution for uniform velocity profiles in feed streams}

\begin{figure}
	\centerline{\includegraphics[scale = .8]{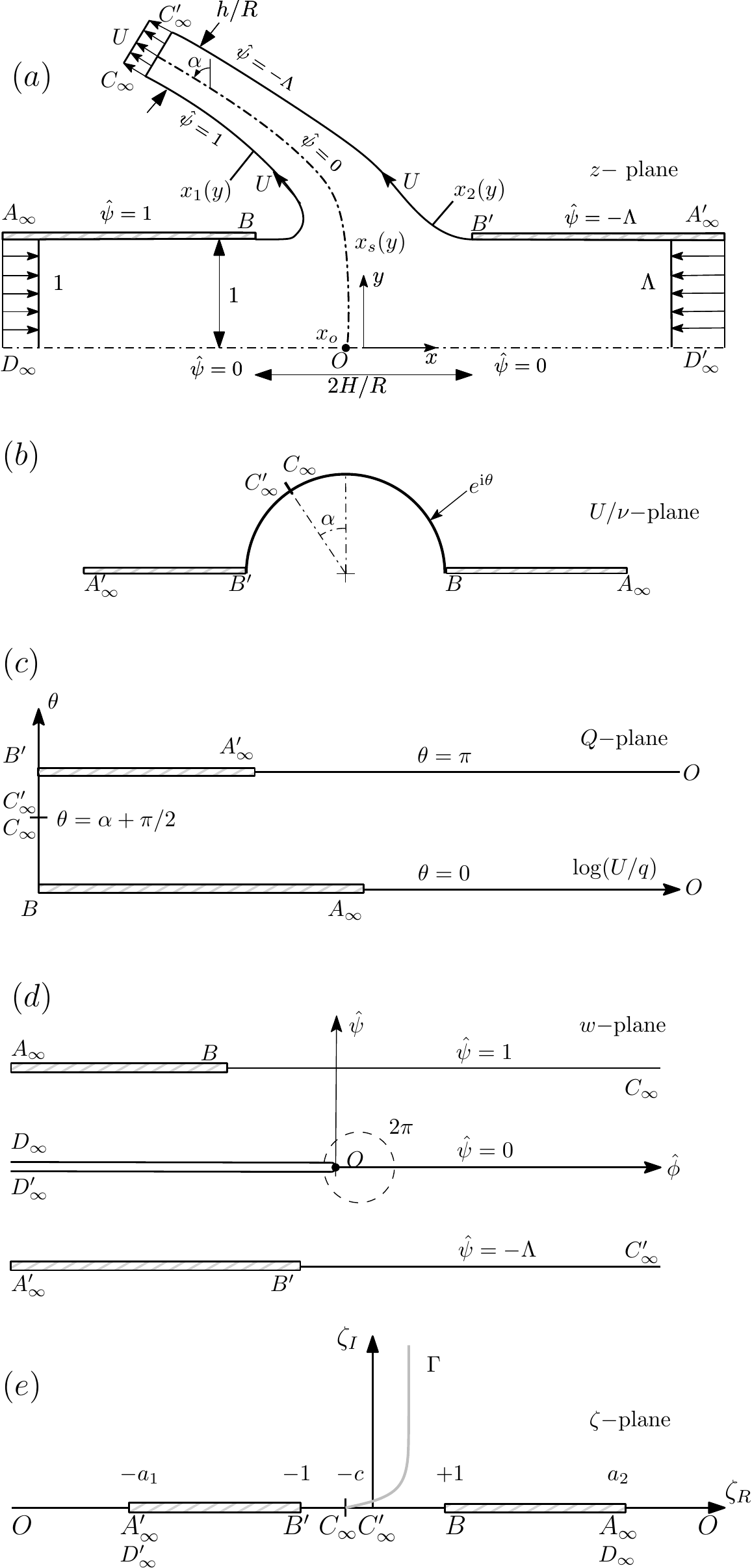}}
	\caption{All relevant complex planes involved in the conformal transformation: (a) $z-$plane, (b) hodographic $\Us/\nu-$plane, (c) $Q-$plane, (d) $w-$plane, (e) upper-half $\zeta-$plane. The contour $\Gamma$ in the $\zeta-$plane represents the separating streamline $\ph = 0$.}	 
	\label{fig:AllComplexPlanes}
\end{figure}

This appendix is concerned with the solution of irrotational counterflows in the general case $H/R \sim 1$, represented schematically in figure \ref{fig:AllComplexPlanes}$a$. The resulting flow can be described in terms of a complex potential 
\beq
w(z) = \hat\phi + \ii \ph,
\eeq
with associated complex velocity 
\beq
\nu(z) = \frac{\dd w}{\dd z} = qe^{-\ii \theta},
\eeq
where $q$ and $\theta$ are the components of the velocity in polar form. The needed conformal transformation, presented below, is based on Kirchhoff's method, delineated in section 11.50 of \citet{thomson1968theoretical}.


\subsection{Definition of conformal planes and associated mapping}

The boundary points employed in the conformal transformation are indicated in figure \ref{fig:AllComplexPlanes}$a$, which includes four additional panels representing the different mapping planes involved in the solution. The points $A_\infty$ and $A_\infty^\prime$ are considered to be at $z=\mp \infty$ respectively. The location of the stagnation point $O$, which generally does not coincide with the origin $z=0$ except in the symmetric case $\Lambda = 1$, must be determined as part of the solution. The speed is constant and equal to $\Us$ along the free streamlines $\bc$ and $\bpcp$ bounding the resulting free jet that forms downstream from the impinging region. This lateral jet becomes uniform at $\ccp$ with width $h/R$ and direction $\alpha$ measured relative to the $y$-axis. 

The objective of the following development is to determine, for given values of $\Lambda$ and $H/R$, (i) the location of the free streamlines $x=x_1(y)$ and $x=x_2(y)$, (ii) the speed $\Us$ along these streamlines, (iii) the width $h/R$ and angle $\alpha$ of the emerging jet, (iv) the location of the stagnation point $z = x_o$, (v) the value $A_o$ of the strain rate at the stagnation point, (vi) the shape of the separating interface $x=x_s(y)$, including its curvature $1/r_c$ at $z = x_o$, and (vii) the parameter $A_o''$ measuring the initial variation of the strain rate along $x=x_s(y)$.

Defining the ratio $\Us/\nu$ we obtain the hodographic plane shown in figure \ref{fig:AllComplexPlanes}$b$. Taking the logarithm 
\beq
Q = \ln(\Us/\nu) = \ln(\Us/q) + \ii \theta \label{defofQ}
\eeq
gives a polygon whose interior may be mapped onto the upper-half $\zeta-$plane using the elementary transformation $\zeta = \cosh Q$, such that
\beq
\zeta = \frac{1}{2}\bigg(\frac{\Us}{q}e^{\ii \theta} + \frac{q}{\Us}e^{-\ii \theta}\bigg),
\eeq
thus defining the location of the different points in the $\zeta-$plane  in terms of the unknown parameter $\Us$. For example, the point $A_\infty$, corresponding to $q=1$ and $\theta = 0$, is mapped onto  $\zeta = (\Us^2+1)/2\Us$. The $\zeta-$plane is shown in figure \ref{fig:AllComplexPlanes}$e$ where the positive constants 
\beq
a_1 = \frac{\Us^2 +\Lambda^2}{2\Lambda \Us}, \qquad a_2 = \frac{\Us^2+1}{2\Us}, \qquad c = \sin\alpha,  \label{a1eq}
\eeq
are the images of $A_\infty^\prime$, $A_\infty$, and $C$, respectively. The points $B$ and $B^\prime$ are mapped onto $\pm1$, while the stagnation point $O$ is mapped to $\zeta=\infty$, as shown in the figure.

Taking $\hat \phi = 0$ at the stagnation point $O$, implies that both $A_\infty D_\infty$ and $A_\infty^\prime D_\infty ^\prime$ correspond to $\hat\phi = -\infty$, while $\ccp$ corresponds to $\hat\phi = \infty$. These selections are accounted for when plotting the $w$ plane in figure \ref{fig:AllComplexPlanes}$d$. 

The next step involves mapping the $w-$plane, regarded as interior to a polygon, onto the upper half $\zeta-$plane by use of the Schwarz-Christoffel transformation. The images in the $\zeta$-plane have already been determined from the mapping $Q \mapsto \zeta$ defined above. Hence, the required transformation is given by 
\begin{align}
\frac{\dd w}{\dd \zeta}= K(\zeta + a_1)^{-1}(\zeta + c)^{-1}(\zeta -a2)^{-1}= K\bigg\{ \frac{\delta}{\zeta-a_2} + \frac{\beta}{\zeta+a_1} - \frac{\gamma}{\zeta + c}\bigg\}, \label{schwarzb}
\end{align} 
where $K=K_R + \ii K_I$ is in general a complex constant and $\delta, \beta, $ and $\gamma$ are given by 
\beq
\delta = \frac{1}{(a_1+a_2)(c+a_2)}, \qquad \beta = \frac{1}{(a_1+a_2)(a_1-c)}, \qquad \gamma =\frac{1}{(a_1-c)(c+a_2)} , \label{deltaeq}
\eeq
obtained from partial fraction expansion, with $\delta + \beta - \gamma = 0$. Integrating~\eqref{schwarzb} yields
\beq
w = K \bigg\{\delta \ln(\zeta - a_2) + \beta \ln(\zeta + a_1) - \gamma \ln(\zeta + c)\bigg\}+L,\label{wofzeta}
\eeq
where $L$ is a complex constant of integration.
Rearranging \eqref{wofzeta} in the form
\beq
\begin{aligned}
	w = \frac{K}{(a_1-c)(c+a_2)(a_1+a_2)}\bigg\{-a_1\ln\bigg(\frac{\zeta+c}{\zeta-a_2}\bigg)-a_2\ln\bigg(\frac{\zeta+c}{\zeta+a_1}\bigg) \\
	+ c\ln\bigg(\frac{\zeta+a_1}{\zeta-a_2}\bigg)     \bigg\} + L, \label{longw}
\end{aligned}
\eeq
and enforcing the condition that the stagnation point $O$ ($w = 0$) is mapped to $\zeta = \infty$, immediately gives $L=0$ since the logarithms in \eqref{longw} vanish as $\zeta \to \infty$. 

On $D_\infty O$, $\ph = 0$ and hence $w$ is real. Since in the $\zeta$-plane the portion $D_\infty O$ corresponds to $\zeta -a_2 >0$, $\zeta + a_1 >0 $, $\zeta + c>0$, all logarithms in~\eqref{wofzeta} are necessarily real. Hence, evaluating \eqref{wofzeta} along $D_\infty O$, where $\hat \phi \ne 0$ and $\ph=0$, reveals that $K_I = 0$, so that $K = K_R$ is purely real. 

Evaluating \eqref{wofzeta} along $\bc$, corresponding to the streamline $\ph = 1$, and equating real and imaginary parts yields 
\beq
K = \frac{(a_1+a_2)(c+a_2)}{\upi}. \label{K1}
\eeq
The constant $K$ may be related to the momentum--flux parameter $\Lambda$, by evaluating \eqref{wofzeta} along  $\bpcp$ for which $\ph = - \Lambda$. Equating real and imaginary parts and solving for $K$ gives 
\beq
K = \frac{\Lambda(a_1-c)(a_1+a_2)}{\upi}. \label{K2}
\eeq
Equating~\eqref{K1} and~\eqref{K2} yields $c = (\Lambda - 1)/2\Us$, relating $c = \sin\alpha$ with $\Lambda$ and $\Us$. Note that this last result may be derived directly from the continuity and momentum integral balances~\eqref{balances}. 

In summary, we have obtained the following mappings to the upper half $\zeta-$plane
\begin{align}
Q(\zeta) &= \cosh^{-1}\zeta = \ln(\zeta + \sqrt{\zeta^2-1}), \label{Qofzeta}\\
w(\zeta) &= K \bigg\{\delta \ln(\zeta - a_2) + \beta \ln(\zeta + a_1) - \gamma \ln(\zeta + c)\bigg\}. \label{wofzeta2}
\end{align}
Equation \eqref{Qofzeta}, together with the definition of $Q$ given in~\eqref{defofQ} provide
\beq
\nu(\zeta) = \frac{\Us}{\zeta + \sqrt{\zeta^2-1}}. \label{nuofzeta}
\eeq 
for the complex velocity.

\subsection{The jet outer boundaries}

In order to compute streamlines, the line element 
\beq
\dd z = \frac{\dd z}{\dd w}\frac{\dd w}{\dd \zeta}\dd \zeta = \frac{1}{\nu}\frac{\dd w}{\dd \zeta}\dd \zeta, \label{lineelement}
\eeq
is expressed as
\beq
\frac{\dd z}{\dd \zeta} = Z(\zeta)=\frac{K}{\Us} \frac{ \zeta + \sqrt{\zeta^2-1}}{(\zeta + a_1)(\zeta+c)(\zeta-a_2)} \label{dzofzeta}
\eeq
with use of~\eqref{schwarzb} and~\eqref{nuofzeta}. Since the free streamlines separating the jets from the ambient gas are mapped onto the segment $-1<\zeta<1$ on the real axis of the $\zeta-$ plane, we may write $\sqrt{\zeta^2 -1} = \ii \sqrt{1-\zeta^2}$. Separating real and imaginary parts of \eqref{dzofzeta} then gives  
\beq
\begin{split}
	\frac{\dd x}{\dd \zeta} &= X(\zeta)=\frac{K}{\Us} \frac{ \zeta }{(\zeta + a_1)(\zeta+c)(\zeta-a_2)} , \label{dxofzeta}\\
	\frac{\dd y}{\dd \zeta} &= Y(\zeta)=\frac{K}{\Us}\frac{\sqrt{1-\zeta^2}}{(\zeta + a_1)(\zeta+c)(\zeta-a_2)}, 
\end{split}
\eeq
which can be integrated to give
\beq
x_1(\zeta) = -H/R + \int_{1}^{\zeta} X(\zeta^\prime) \dd \zeta^\prime \qquad \text{and} \qquad
y(\zeta) = 1 + \int_{1}^{\zeta} Y(\zeta^\prime) \dd \zeta^\prime\\ \label{integralformoffreestreamlines1}
\eeq
as an implicit representation for $x_1(y)$, and
\beq
x_2(\zeta) = H/R + \int_{-1}^{\zeta} X(\zeta^\prime) \dd \zeta^\prime \qquad \text{and} \qquad
y(\zeta) = 1 + \int_{-1}^{\zeta} Y(\zeta^\prime) \dd \zeta^\prime\\ \label{integralformoffreestreamlines2}
\eeq
as an implicit representation for $x_2(y)$. 

\subsection{The emerging jet}

The two streamlines $x_1(y)$ and $x_2(y)$ become parallel as $y \rightarrow \infty$. The thickness of the resulting jet, given by 
\beq
h/R= \cos\alpha \lim_{y \rightarrow \infty} (x_2-x_1),
\eeq
can be obtained by evaluating the first equations in \eqref{integralformoffreestreamlines1} and \eqref{integralformoffreestreamlines2} as $\zeta \to -c$ and taking the difference to yield
\begin{align}
h/R =\cos\alpha \bigg\{2H/R + \frac{1}{\upi \Us^2}\bigg[&(U^2+1)\ln\bigg(\frac{U -1}{U+1}\bigg) + (U^2+\Lambda^2) \ln\bigg(\frac{U -\Lambda}{U+\Lambda}\bigg) \nonumber \\
&+ \frac{\Lambda^2-1}{2}\ln\bigg(\frac{2U+\Lambda-1}{2U-\Lambda+1}\bigg)      \bigg] \bigg\}, \label{deltax2}
\end{align}
after simplification with use made of~\eqref{a1eq}, \eqref{deltaeq} and~\eqref{K1}. This equation together with the continuity and momentum integral balances~\eqref{balances} 
can be used to determine the three unknowns $\alpha$, $h/R$, and $\Us$ in terms of the parameters $\Lambda$ and $H/R$. The value of $\Us$ can then be used in~\eqref{K1} to compute the factor
\beq
K=\frac{(U^2+\Lambda)^2(\Lambda+1)}{4\Lambda U^2}
\eeq
for the functions $Z(\zeta)$,  $X(\zeta)$, and $Y(\zeta)$ defined in~\eqref{dzofzeta} and \eqref{dxofzeta}. 

\subsection{The interface separating the jets}

The stagnation point $\boldsymbol{x}_o=(x_o,0)$ is located at $\zeta = \infty$ in the $\zeta-$plane. Integrating \eqref{dzofzeta} along $BA_\infty O$ on the real axis of the $\zeta-$ plane, for which $\zeta>1$ so that $\dd z = \dd x$, gives 
\beq
x_o = -H/R+ \dashint_1^\infty Z(\zeta^\prime)\dd \zeta^\prime, \label{xstageqn}
\eeq
with the principal value of the integral needed to account for the singularity at $\zeta = a_2$. Equation \eqref{xstageqn} may be integrated numerically to obtain $x_o$.

The separating streamline departs from $(x,y)=(x_o,0)$. Its shape can be computed by integrating \eqref{dzofzeta} along the corresponding contour $\ph = 0$ in the $\zeta-$plane. This contour, denoted by $\Gamma$ in figure \ref{fig:AllComplexPlanes}$e$, is obtained from the condition
\beq
\ph(\zeta) = \Imag[w(\zeta)] = 0, \label{contour}
\eeq
with $w(\zeta)$ defined by the transformation given in \eqref{wofzeta2}. The constant of integration is determined by imposing the correspondence of the points $\zeta = \infty$ and $z = x_o$. 

\subsection{Flow properties near the stagnation point}

As discussed in the main text, the structure of the mixing layer near the stagnation point, of particular interest in counterflow--flame studies, depends on the local strain rate $A_o=A_1=(\rho_2/\rho_1)^{1/2}A_2$, with departures from selfsimilarity measured by the local values of the mixing-layer curvature $r_c$ and of the parameter $A_o''=A''_1=(\rho_2/\rho_1)^{1/2}A''_2$, the latter entering in the velocity distributions on both sides of the mixing layer given in~\eqref{veldists}. These quantities can be evaluated by writing~\eqref{AorcAo''} in terms of the complex potential to give
\beq
A_o = - \Real\bigg[\frac{\dd^2 w}{\dd z^2}\bigg]_{z=x_o},  \quad \frac{1}{r_c}=-\frac{1}{3A_o} \Real\bigg[\frac{\dd^3 w}{\dd z^3}\bigg]_{z=x_o}, \; \text{and} \; A''_o = \Real\bigg[\frac{\dd^4 w}{\dd z^4}\bigg]_{z=x_o}. \label{AoLambda_old}
\eeq
Since an explicit expression for $w = w(z)$ is not available, in evaluating the above equations it is convenient to use the expression \eqref{wofzeta2} for $w = w(\zeta)$ and use the chain rule to express the derivatives in \eqref{AoLambda_old} in the form $\dd/\dd z=Z(\zeta)^{-1} \dd/\dd \zeta$, where $Z(\zeta)={\rm d} z/{\rm d}\zeta$ is defined in~\eqref{dzofzeta}. Evaluating the derivatives as $\zeta \to \infty$, corresponding to the location of the stagnation point $z=x_o$ on the $\zeta$-plane, provides
\beq
\begin{aligned}
	& A_o=\frac{\upi \Lambda U^4}{(U^2+\Lambda)^2(\Lambda+1)}, \; \frac{1}{r_c}=\frac{\upi U^2 (U^2-2 \Lambda)(\Lambda-1)}{3 (\Lambda+1)(U^2+\Lambda)^2}, \; A_o''=\frac{\upi^3 U^8 \Lambda}{(\Lambda+1)^3 (U^2+\Lambda)^6} \\
	&\times [4U^2\Lambda (\Lambda-2) (2\Lambda-1) -U^4 (\Lambda^2-4\Lambda+1)-2 \Lambda^2 (3\Lambda^2-7\Lambda+3)],
\end{aligned}
\label{AoLambda}
\eeq
after simplifying the result.

\bibliographystyle{jfm}
\bibliography{references}

\end{document}